\documentclass[12pt]{article}
\usepackage{amsmath,epsfig,graphicx,verbatim}
\usepackage{color}

\include{epsf} 
\newcommand\as{\alpha_{\mathrm{S}}} 
\newcommand\f[2]{\frac{#1}{#2}} 
\def\beq{\begin{equation}} 
\def\eeq{\end{equation}} 
\def\beeq{\begin{eqnarray}} 
\def\eeeq{\end{eqnarray}} 
 
\def\to{\rightarrow}
 
\def\nn{\nonumber} 
 
\def\tL{{\tilde L}}
 
\def\qt{q_T}
\def\mures{\mu_{res}}

\def\mgg{M_{\gamma \gamma}}
\def\Mgg{M_{\gamma \gamma}}
\def\gg{\gamma \gamma}
\def\Dpgg{\Delta \Phi_{\gamma\gamma} }

\begin{document} 
\begin{titlepage}
\renewcommand{\thefootnote}{\fnsymbol{footnote}}
\vspace*{2cm}

\begin{center}
{\Large \bf Diphoton production at hadron colliders:\\[0.2cm]
transverse-momentum resummation \\[0.4cm] 
at next-to-next-to-leading logarithmic accuracy}
\end{center}

\par \vspace{2mm}
\begin{center}
{\bf Leandro Cieri${}^{(a)}$,}
{\bf Francesco Coradeschi${}^{(b)}$, and 
Daniel de Florian${}^{(c)}$}\\
\vspace{5mm}
${}^{(a)}$ Dipartimento di Fisica, Universit\`a di Roma ``La Sapienza'' and\\
INFN, Sezione di Roma, I-00185 Rome, Italy\\
${}^{(b)}$INFN, Sezione di Firenze and
Dipartimento di Fisica, Universit\`a di Firenze,\\
I-50019 Sesto Fiorentino, Florence, Italy\\
${}^{(c)}$Departamento de F\'\i sica, FCEYN, Universidad de Buenos Aires,\\
(1428) Pabell\'on 1 Ciudad Universitaria, Capital Federal, Argentina\\
\vspace{25mm}
\end{center}

\par \vspace{2mm}
\begin{center} {\large \bf Abstract} \end{center}
\begin{quote}
\pretolerance 10000

We consider the transverse-momentum ($q_T$)
distribution of a diphoton pair produced
in hadron collisions. 
At small values of $q_T$, we resum the logarithmically-enhanced 
perturbative QCD contributions
up to next-to-next-to-leading logarithmic accuracy. 
At intermediate and large values
of $q_T$,
we consistently combine resummation with the 
known next-to-leading order
perturbative result.
All perturbative terms up to order $\as^2$ are included in our computation 
which, after integration over $q_T$, reproduces
the known next-to-next-to-leading order result for the diphoton pair production 
total cross section. 
We present a comparison with LHC data and an estimate of the perturbative 
accuracy of the theoretical calculation by performing the corresponding 
variation of scales. In general we observe that the effect of the resummation
is not only to recover the predictivity of the calculation at small transverse momentum,
but also to improve substantially the agreement with the experimental data.

\end{quote}

\vspace*{\fill}
\begin{flushleft}
May 2015

\end{flushleft}
\end{titlepage}

\setcounter{footnote}{1}
\renewcommand{\thefootnote}{\fnsymbol{footnote}}
\section{Introduction}

Diphoton production at hadronic colliders is a very relevant process, both from the point of view of
testing the Standard Model (SM) predictions~\cite{Chatrchyan:2011qt,Aaltonen:2011vk,Chatrchyan:2014fsa,Abazov:2013pua,Aad:2011mh,Aad:2012tba} as for new physics searches.

While a general understanding of QCD processes at hadronic collisions poses a serious challenge due the complicated environment, 
 \emph{direct} or \emph{prompt} photons provide an ideal probe  since they constitute a theoretically and experimentally
\emph{clean} final state. From the theory side, because they do not have QCD interactions with other final state particles and, 
from the experimental side, because their energies and momenta can be measured with high precision in modern electromagnetic calorimeters.

Besides purely QCD-related considerations, diphoton final states have played a crucial role in the recent discovery of a new boson at
the LHC \cite{cha:2012gu,aad:2012gk}, whose properties are compatible with those of the SM Higgs. They are also important in many
new physics scenarios \cite{:2012afa,:2012mx}, in particular in the search for extra-dimensions~\cite{Aad:2012cy} or supersymmetry~\cite{CMS:2012un}. 

In this paper we are interested in the process $pp\rightarrow \gamma \gamma X$, and in particular in the transverse-momentum ($q_T$) spectrum of the diphoton pair.
The lowest-order process ($\mathcal{O}(\alpha_S^0)$) occurs \textit{via} the quark annihilation subprocess $q\bar{q}\rightarrow \gamma\gamma$.
The QCD corrections at the first order in the strong coupling $\alpha_S$  involve 
quark annihilation  and a new partonic channel, \textit{via} the subprocess $qg \rightarrow \gamma \gamma q$.
First order corrections have been computed and implemented in several fully-differential Monte Carlo codes \cite{Binoth:1999qq,Bern:2002jx,Campbell:2011bn,Balazs:2007hr}.
At the second order in the strong coupling $\alpha_S$ the $gg$ channel starts to contribute, and the large gluon--gluon luminosity makes
this channel potentially sizeable.

The amplitudes needed to evaluate the corrections at the second order in the strong coupling $\alpha_S$, 
for diphoton production, have been presented
in \cite{Dicus:1987fk,Barger:1989yd,Bern:1994fz,Anastasiou:2002zn}, and first
put together in a complete and consistent $\mathcal{O}(\alpha_S^2)$ calculation in the \texttt{2$\gamma$NNLO} code \cite{Catani:2011qz}.
The next-order gluonic corrections to the {\it box contribution} (which are part of the N$^3$LO QCD corrections to diphoton production)
were also computed in ref.~\cite{Bern:2002jx} and found to have a moderate quantitative effect.

The calculation of the $q_T$ spectrum poses an additional challenge with respect to more inclusive calculations, such as the total
cross section. In the large-$q_T$ region ($q_T\sim \Mgg$), where the transverse momentum is of the order of the diphoton invariant
mass $\Mgg$, calculations based on the truncation of the perturbative series at a fixed order in $\as$  are theoretically justified.
In this region, the QCD radiative corrections are known up to the next-to-leading order (NLO), including the corresponding partonic
scattering amplitudes with $X=2$~partons (at the tree level \cite{Barger:1989yd}) and the partonic scattering amplitudes with 
$X=1$~parton (up to the one-loop level \cite{Bern:1994fz}). We remind the reader that at least one additional parton is needed in
order to have $q_T \neq 0$ for the diphoton pair. The $q_T$ spectrun of the diphoton pair has been calculated in fully-differential 
Monte Carlo codes at LO~\cite{Binoth:1999qq,Bern:2002jx,Campbell:2011bn,Balazs:2007hr} and at NLO~\cite{Catani:2011qz,DelDuca:2003uz,Gehrmann:2013aga}. Recently, first calculations for diphoton production in association with two~\cite{Bern:2014vza,Gehrmann:2013bga,Badger:2013ava} and three~\cite{Badger:2013ava} jets at NLO became available.

The bulk of the diphoton events is produced in the small-$q_T$ region ($q_T\ll \mgg$), where the convergence of the
fixed-order expansion is spoiled by the presence of large logarithmic terms, $\as^n\ln^m (\mgg^2/q_T^2)$.
In order to obtain reliable predictions these logarithmically-enhanced terms have to be systematically resummed to all perturbative orders
\cite{Dokshitzer:hw}--\cite{Catani:2013tia}. The resummed calculation, valid at small values of $q_T$, and the fixed-order one at large
$q_T$ have then to be consistently matched to obtain a pQCD prediction for the entire range of transverse momenta.

We use the transverse-momentum resummation formalism proposed  in Refs.~\cite{Catani:2000vq,Bozzi:2005wk,Bozzi:2007pn}
(see also \cite{Catani:2010pd} for processes initiated by $gg$ annihilation). 
The formalism is valid for a generic process in which a high-mass system of non strongly-interacting particles is produced 
in hadron-hadron collisions.
The method has so far been applied to the production of the Standard Model (SM) Higgs boson 
\cite{ Bozzi:2005wk, Bozzi:2007pn, Bozzi:2003jy, deFlorian:2011xf,deFlorian:2012mx}, Higgs boson production in bottom quark annihilation~\cite{Harlander:2014hya}, Higgs boson production via gluon fusion in the MSSM~\cite{Harlander:2014uea}, single vector bosons at NLL+LO \cite{Bozzi:2008bb} and at NNLL+NLO \cite{Bozzi:2010xn},
$WW$ \cite{Grazzini:2005vw,Meade:2014fca} and $ZZ$ \cite{Frederix:2008vb} pairs, slepton pairs \cite{Bozzi:2006fw}, and DY lepton pairs in polarized collisions \cite{Jiro}.

Finally, note that besides the direct photon production from the hard subprocess, photons can also
arise from the fragmentation of QCD partons. The computation of fragmentation subprocesses requires (the poorly known)
non-perturbative information, in the form of parton fragmentation functions of the photon (the complete single- and
double-fragmentation contributions are implemented in \texttt{DIPHOX} \cite{Binoth:1999qq} for diphoton production at the first order in $\as$).
However, the effect of the fragmentation
contributions is sizeably reduced by the \emph{photon isolation} criteria that are necessarily
applied in hadron collider experiments to suppress the very large irreducible background (\textit{e.g.}, photons that are faked by jets
or produced by hadron decays). Two such criteria are the so-called ``standard'' cone isolation and the ``smooth''
cone isolation proposed by Frixione \cite{Frixione:1998jh}. The standard cone isolation is easily implemented in experiments,
but it only suppresses a fraction of the fragmentation contribution.
By contrast, the smooth cone isolation (formally) eliminates the entire fragmentation contribution. 
For all of the results presented in this paper we rely on the smooth isolation prescription, which, for the parameters used
in the experimental analysis reproduces the standard result within a $1\%$ accuracy~\cite{Butterworth:2014efa}.

The paper is organized as follows. In Sect.~\ref{sec:theory} we briefly review the resummation formalism of Refs.
\cite{Catani:2000vq,Bozzi:2005wk,Bozzi:2007pn}.
In Sect.~\ref{sec:results} we present numerical results and we comment on their comparison with the LHC data \cite{Aad:2012tba}.
We also study the scale dependence of our results with the purpose of estimating the corresponding perturbative uncertainty. 
In Sect.~\ref{sec:summa} we summarize our results.

\section{Transverse-momentum resummation}
\label{sec:theory}

We briefly recall the main points of the transverse-momentum resummation formalism of Refs.~
\cite{Catani:2000vq,Bozzi:2005wk,Bozzi:2007pn}, referring to the original papers for the full details.
The formalism is general, as long as the measured final state is composed of non strongly-interacting particles (transverse-momentum resummation for strongly-interacting final states, such as heavy-quark production, has been developed in Refs.~\cite{Li:2013mia,Catani:2014qha}). Here we specialize
to the case of diphoton production only for ease of reading. The inclusive hard-scattering process considered is
\begin{equation}
h_1(p_1) + h_2(p_2) \;\to\; \gamma \gamma(\Mgg,q_T,y) + X \;\;,   
\label{first}
\end{equation}
where $h_1$ and $h_2$ are the colliding hadrons with momenta $p_1$ and $p_2$, $\gamma \gamma$ is the diphoton pair
with invariant mass $\Mgg$, transverse momentum $q_T$ and rapidity $y$, and $X$ is an arbitrary and undetected final state. 

The corresponding fully differential cross section, in $q_T$, $\Mgg$ and $y$, which we denote for simplicity (since our focus is on the $q_T$
distribution) by $d\sigma_{\gg}/dq_T^2$, can be written using the factorization formula as
\begin{equation}
\label{dcross}
\f{d\sigma_{\gg}}{d q_T^2}(q_T,\Mgg,s)= \sum_{a,b}
\int_0^1 \!\!\!dx_1 \,\int_0^1 \!\!\!dx_2 \,f_{a/h_1}(x_1,\mu_F^2)
\,f_{b/h_2}(x_2,\mu_F^2) \;
\f{d{\hat \sigma}^{\gg}_{ab}}{d q_T^2}(q_T, \Mgg,y,{\hat s};
\as,\mu_R^2,\mu_F^2)
\end{equation}
(up to power-suppressed corrections), where the $f_{a/h}(x,\mu_F^2)$ ($a=q,{\bar q}, g$)
are the parton densities of the hadron $h$ at the factorization scale $\mu_F$, $\as \equiv \as(\mu_R^2)$,
$d\hat\sigma^{\gg}_{ab}/d{q_T^2}$ is the pQCD \emph{partonic cross section}, 
$s$ ($\hat s = x_1 x_2 s$) 
is the square of the hadronic (partonic) centre--of--mass  energy,  and $\mu_R$ is the renormalization scale. 

In the region where $q_T \sim  \Mgg$ the QCD perturbative series is controlled by a small expansion parameter, 
$\as(\Mgg)$, and a fixed-order calculation of the partonic cross section is theoretically justified. In this region, 
the QCD radiative corrections are known up to next-to-leading order (NLO)~\cite{Dicus:1987fk,Barger:1989yd,Bern:1994fz,Anastasiou:2002zn}. 

In the small-$q_T$ region ($q_T\ll \Mgg$),
the convergence of the fixed-order perturbative expansion is spoiled by the presence
of powers of large logarithmic terms,  $\as^n\ln^m (\Mgg^2/q_T^2)$.
To obtain reliable predictions these terms have to be resummed to all orders.

To perform the resummation, we start by decomposing the partonic cross section as
\begin{equation}
\label{resplusfin}
\f{d{\hat \sigma}^{\gg}_{ab}}{dq_T^2}=
\f{d{\hat \sigma}_{\gg\,ab}^{(\rm res.)}}{dq_T^2}
+\f{d{\hat \sigma}_{\gg\,ab}^{(\rm fin.)}}{dq_T^2}\; .
\end{equation}
The first term on the right-hand side contains all the logarithmically-enhanced contributions,
which have to be resummed to all orders in $\as$,
while the second term
is free of such contributions and can thus be evaluated at fixed order in perturbation theory. 
Using the Fourier transformation between the conjugate variables 
$q_T$ and $b$ ($b$ is the impact parameter),
the resummed component $d{\hat \sigma}^{({\rm res.})}_{\gg\,ab}$
can be expressed as
\begin{equation}
\label{resum}
\f{d{\hat \sigma}_{\gg \,ab}^{(\rm res.)}}{dq_T^2}(q_T,\Mgg,y,{\hat s};
\as,\mu_R^2,\mu_F^2) 
=\f{\Mgg^2}{\hat s} \;
\int_0^\infty db \; \f{b}{2} \;J_0(b q_T) 
\;{\cal W}_{ab}^{\gg}(b,\Mgg,y,{\hat s};\as,\mu_R^2,\mu_F^2) \;,
\end{equation}
where $J_0(x)$ is the $0$th-order Bessel function.
The form factor ${\cal W}^{\gg}$ is best expressed in terms of its \emph{double} Mellin moments ${\cal W}_{N_1 N_2}^{\gg}$, taken with
respect to the variables $z_1, \ z_2$ at fixed $\Mgg$, with
\begin{equation}
z_1 z_2 \equiv z = \frac{\Mgg^2}{\hat{s}}, \quad \frac{z_1}{z_2} = e^{2 y};
\end{equation}
the resummation structure of ${\cal W}_{N_1 N_2}^{\gg}$ can be organized in an exponential 
form~\footnote{For the sake of simplicity we consider here only
the case of  the diagonal terms in the flavour space 
of the partonic indices $a,b$. For a detailed discussion, we refer to Ref.~\cite{Bozzi:2005wk,Bozzi:2007pn}.}
\begin{align}
\label{wtilde}
{\cal W}_{N_1 N_2}^{\gg}(b,\Mgg,y;\as,\mu_R^2,\mu_F^2)
& ={\cal H}_{N_1 N_2}^{\gg}\left(\Mgg, 
\as;\Mgg^2/\mu^2_R,\Mgg^2/\mu^2_F,\Mgg^2/\mures^2
\right) \nonumber \\
&\times \exp\{{\cal G}_{N_1 N_2}(\as,L;\Mgg^2/\mu^2_R,\Mgg^2/\mures^2
)\}
\;\;,
\end{align}
were we have defined the logarithmic expansion parameter $L\equiv \ln ({\mures^2 b^2}/{b_0^2})$,
and $b_0=2e^{-\gamma_E}$ ($\gamma_E=0.5772...$ 
is the Euler number).
The scale $\mures$ ($\mures\sim \Mgg$), 
which appears on the right-hand side of Eq.~(\ref{wtilde}),
is the resummation scale \cite{Bozzi:2005wk}. Although ${\cal W}_{N_1 N_2}^{\gg}$ (i.e., the product
${\cal H}_{N_1 N_2}^{\gg} \times \exp\{{\cal G}_{N_1 N_2}\}$) does not depend on $\mures$ when
evaluated to all perturbative orders, its explicit dependence on $\mures$
appears when ${\cal W}_{N_1 N_2}^{\gg}$ is computed by truncation of the resummed
expression at some level of logarithmic accuracy (see Eq.~(\ref{exponent})
below). 
Variations of $\mures$ around $\Mgg$ can thus be used to estimate the size of yet uncalculated 
higher-order logarithmic contributions.

The form factor $\exp\{ {\cal G}_{N_1 N_2}\}$ is universal\footnote{The form factor does not depend on the final state;
all the hard-scattering processes initiated by $q\bar{q}$ ($gg$) annihilation have the same form factor.} and
contains all the terms that order-by-order in $\as$ are logarithmically divergent 
as $b \to \infty$ (or, equivalently, $q_T\to 0$). The resummed logarithmic expansion of the exponent ${\cal G}_{N_1 N_2}$ 
is defined as follows:
\begin{align}
\label{exponent}
{\cal G}_{N_1 N_2}(\as, L;\Mgg^2/\mu^2_R,\Mgg^2/\mures^2)&=L \;g^{(1)}(\as L)+g_{N_1 N_2}^{(2)}(\as L;\Mgg^2/\mu_R^2,\Mgg^2/\mures^2)\nn\\
+ & \f{\as}{\pi} g_{N_1 N_2}^{(3)}(\as L,\Mgg^2/\mu_R^2,\Mgg^2/\mures^2)+\dots
\end{align}
where the term $L\, g^{(1)}$ collects the leading logarithmic (LL) $\mathcal{O}(\alpha_s^{p+n}L^{n+1})$
contributions, the function $g_{N_1 N_2}^{(2)}$ includes
the next-to-leading leading logarithmic (NLL) $\mathcal{O}(\alpha_s^{p+n}L^{n})$ contributions \cite{Kodaira:1981nh}, 
$g_{N_1 N_2}^{(3)}$ controls the NNLL $\mathcal{O}(\alpha_s^{p+n}L^{n-1})$ terms \cite{Davies:1984hs, Davies:1984sp, deFlorian:2000pr,Becher:2010tm}
and so forth; $p$ is the number of powers of $\alpha_s$ in the LO (Born) process.
In Eq.~\eqref{exponent}, $\as L$ is formally of order $1$, so there is
an explicit $\mathcal{O}(\as)$ suppression between different logarithmic orders. The explicit form of the functions
$g^{(1)}$, $g_{N_1 N_2}^{(2)}$ and $g_{N_1 N_2}^{(3)}$ can be found in Ref.~\cite{Bozzi:2005wk}.
The process dependent function ${\cal H}_{N_1 N_2}^{\gg}$ 
does not depend on the impact parameter $b$ and it 
includes all the perturbative
terms that behave as constants as $b \to \infty$. 
It can thus be expanded in powers of $\as$:
\begin{align}
\label{hexpan}
{\cal H}_{N_1 N_2}^{\gg}(\Mgg,\as;\Mgg^2/\mu^2_R,\Mgg^2/\mu^2_F,\Mgg^2/\mures^2)&=
\sigma_{\gg}^{(0)}(\alpha_s, \Mgg)
\Bigl[ 1+ \f{\as}{\pi} \,{\cal H}_{N_1 N_2}^{\gg \,(1)}(\Mgg^2/\mu^2_F,\Mgg^2/\mures^2) 
\Bigr. \nn \\
+ \Bigl. \left(\f{\as}{\pi}\right)^2 &
\,{\cal H}_{N_1 N_2}^{\gg \,(2)}(\Mgg^2/\mu^2_R,\Mgg^2/\mu^2_F,\Mgg^2/\mures^2)+\dots \Bigr],
\end{align}
where $\sigma_{\gg}^{(0)}$ is the partonic cross section at the Born level. Since
the formalism applies to non strongly-interacting final states, in general the Born cross-section can
only correspond to a $q \bar{q}$ or $gg$ initial state. In the specific case of the diphoton production, both
channels contribute, but at different orders in $\alpha_s$: the $q\bar{q}$ subprocess initiates as a pure QED process
($\mathcal{O}(\alpha_s)^0$), while the $gg$ one requires a fermion loop, starting at $\mathcal{O}(\alpha_s)^2$.

In the present work, we keep contributions up to an uniform order in $\alpha_s$ (and all orders in $\as L$), namely
up to $\as^n L^{n-1}$. For the $q\bar{q}$ channel, this requires the inclusion of the $\mathcal{H}$ coefficients of Eq.~\eqref{hexpan} up to order $2$: 
the first-order coefficients ${\cal H}_{N_1 N_2}^{\gg(1)}$ are known since a long time \cite{deFlorian:2000pr},
while the second-order coefficients ${\cal H}_{N_1 N_2}^{\gg(2)}$ were computed only relatively recently~\cite{Catani:2011qz,Catani:2013tia}.
For the $gg$ channel, it is sufficient to include the leading $\mathcal{H}$ contribution (that is, the Born cross-section) and the
appropriate $\mathcal{G}$ in the exponential of Eq.~\eqref{exponent}. Since it does not require any additional numerical effort,
we decided, in all the plots presented in the paper, to include all the terms up to $g_{N_1 N_2}^{(3)}$ in the exponential
$\mathcal{G}$ factor also for this channel. In this way, we technically include some terms which are of higher order in $\alpha_s$ with respect to
those in the $q\bar{q}$ channel; however we checked that those terms 
result in a negligible numerical effect (at 1\% accuracy), that is, the
difference produced by including the higher order terms is within the error bands 
obtained by the scale variations, which verifies the stability of the calculation.

Within a straightforward (`naive') implementation of Eq.~(\ref{wtilde}), 
the resummation of the large logarithmic contributions 
would affect
not only the small-$q_T$ region, but also the region of large values of $q_T$.
This can easily be understood by observing that the logarithmic expansion 
parameter $L$ 
diverges also
when $b\to 0$.
To reduce the impact of unjustified higher-order contributions in the large-$q_T$ region,
the logarithmic variable $L$ in Eq.~(\ref{wtilde})
is actually replaced 
by $\tL\equiv \ln \left({\mures^2 b^2}/{b_0^2}+1\right)$ 
\cite{Bozzi:2005wk, Bozzi:2003jy}.
This (unitarity related) replacement has an additional and relevant
consequence: after inclusion of the finite component (see Eq.~(\ref{resfin})),  
we exactly recover the fixed-order perturbative value of the total cross section
upon integration of the $q_T$  distribution over $q_T$
(i.e., the resummed terms give a vanishing contribution upon integration over $q_T$).

We now turn to consider
the finite component of the transverse-momentum cross section
(see Eq.~(\ref{resplusfin})).
Since $d\sigma_{\gg}^{({\rm fin.})}$ does not contain large logarithmic terms
in the small-$q_T$ region,
it can be evaluated by truncation of the perturbative series
at a given fixed order.
In practice, the finite component is computed starting from the usual
fixed-order perturbative truncation of the partonic cross section and
subtracting the expansion of the resummed part at the same perturbative order.
Introducing the subscript f.o. to denote the perturbative truncation of the
various terms, we have:
\begin{equation}
\label{resfin}
\Bigl[ \f{d{\hat \sigma}_{\gg \,ab}^{(\rm fin.)}}{d q_T^2} \Bigr]_{\rm f.o.} =
\Bigl[\f{d{\hat \sigma}_{\gg \,ab}^{}}{d q_T^2}\Bigr]_{\rm f.o.}
- \Bigl[ \f{d{\hat \sigma}_{\gg \,ab}^{(\rm res.)}}{d q_T^2}\Bigr]_{\rm f.o.} \;.
\end{equation} 
This matching procedure 
between resummed and finite contributions guarantees
to achieve uniform theoretical accuracy 
over 
the region from small to intermediate values of transverse momenta. 
At large values of $q_T$,
the resummation (and matching) procedure is eventually superseded by the
customary fixed-order calculations 
(their theoretical accuracy in the large-$q_T$ region cannot be
improved by resummation of the logarithmic terms that dominate 
in the small-$q_T$ region).

In summary,
the inclusion of the functions $g^{(1)}$, $g_{N_1 N_2}^{(2)}$,
${\cal H}_{N_1 N_2}^{\gg (1)}$ in the resummed component,
together with the evaluation of the finite component at LO (i.e. at ${\cal O}(\as)$),
allows us to perform the resummation at NLL+LO accuracy.
This is the theoretical accuracy used in previous studies \cite{Balazs:2007hr,Balazs:2006cc,Nadolsky:2002gj,Balazs:1999yf}
of the diphoton $q_T$ distribution.
Including also the functions $g_{N_1 N_2}^{(3)}$ and ${\cal H}_{N_1 N_2}^{\gg(2)}$, together 
with the finite component at NLO (i.e. at ${\cal O}(\as^2)$)
leads to full NNLL+NLO accuracy.

Using the ${\cal H}_{N_1 N_2}^{\gg(2)}$ 
coefficient~\cite{Catani:2011qz,Catani:2013tia},
we are thus able to present the complete result for the
diphoton $q_T$-distribution up to NNLL+NLO accuracy.
We 
point out
that the NNLL+NLO (NLL+LO) result includes the {\em full} NNLO (NLO)
perturbative contribution in the small-$\qt$ region.
In particular,  the NNLO (NLO) result for the total cross section  
is exactly recovered upon integration
over $q_T$ of the differential cross section $d \sigma_{\gg}/dq_T$ at NNLL+NLO
(NLL+LO) accuracy.

We conclude this section with some comments on the numerical implementation
of our calculation.
Within our formalism, the resummation factor ${\cal W}_{N_1 N_2}^{\gg}(b,\Mgg)$
is directly defined, at fixed $\Mgg$,
in the space of the conjugate variables $b$ and $N_1$, $N_2$. 
To obtain the hadronic cross section,
we have to perform inverse integral transformations: the Bessel transformation in Eq.~(\ref{resum}) and an inverse
Mellin transformation, implemented following the prescription introduced in Ref. \cite{Kulesza:2002rh}.
These integrals are carried out numerically. The Mellin inversion 
requires the numerical evaluation of some basic $N$-moment functions that appear in the expression of the  
the second-order coefficients ${\cal H}_{N_1N_2}^{\gg(2)}$
~\cite{Catani:2011qz,Catani:2013tia}: this evaluation 
has to be performed for complex values of $N_1$ and $N_2$; to evaluate some of the needed special function at complex value,
we use the numerical routines of Ref.~\cite{Blumlein:2000hw}.
We recall \cite{Bozzi:2005wk} that the resummed form factor 
$\exp \{{\cal G}_{N_1 N_2}(\as(\mu_R^2),{\widetilde L})\}$
is singular at  the values of $b$ where $\as(\mu_R^2) {\widetilde L} \geq \pi/\beta_0$ 
($\beta_0$ is the first-order coefficient of the QCD $\beta$ function). We avoid this singularity by 
introducing a smooth effective cut-off at small $b$, which is shown to have a negligible effect in the final result.

It is known that at small values of $q_T$, the perturbative QCD approach has to be supplemented with non-perturbative contributions, since they become relevant as $q_T$ decreases. A discussion on non-perturbative effects on the $q_T$ distribution is presented in Ref.~\cite{Bozzi:2005wk}, and related quantitative results are shown in Sect. \ref{sec:results}.

\section{Numerical results for diphoton production at the LHC}
\label{sec:results}

In this section 
we consider diphoton production in $pp$ collisions at LHC
energies ($\sqrt{s}=7$~TeV). We present our resummed results at NNLL+NLO accuracy, and compare them with NLL+LO predictions
and with available LHC data~\cite{Aad:2012tba}.
Since the present formulation of the $q_T$ resummation formalism,
is restricted to the production of colourless systems $F$, it does not 
treat parton fragmentation subprocesses (here $F$ includes one or two coloured
partons that fragment); therefore, we concentrate on the direct production of diphotons, and 
we rely on the smooth cone isolation criterion proposed by Frixione~\cite{Frixione:1998jh} (see also 
Ref.~\cite{Frixione:1999gr,Catani:2000jh}) which is defined by requesting
\begin{eqnarray}\label{Eq:Isol_frixcriterion}     
&\sum E_{T}^{had} \leq E_{T \, max}~\chi(r)\;, \;\;\;\;\nonumber\\
&\mbox{inside any} \;\;      
r^{2}=\left( y - y_{\gamma} \right)^{2} +    
\left(  \phi - \phi_{\gamma} \right)^{2}  \leq R^{2}  \;,    
\end{eqnarray}
with a suitable choice for the function $\chi(r)$. This function has to vanish smoothly when its argument goes to zero ($\chi(r) \rightarrow 0 \;,\; \mbox{if} \;\; r \rightarrow 0\,$), and it has to verify \mbox{$\; 0<\chi(r)< 1$}, if \mbox{$0<r<R\,\,.$} One possible choice is
\begin{equation}
\label{Eq:Isol_chinormal}
\chi(r) = \left( \frac{1-\cos (r)}{1-\cos R} \right)^{n}\;,
\end{equation}
where $n$ is typically chosen as $n=1$. This condition implies that, closer to the photon, less hadronic activity is allowed inside the cone. At $r=0$,
when the parton and the photon are exactly collinear, the energy deposited inside the cone is required to be exactly equal to zero, and the
fragmentation component  (which is a purely collinear phenomenon in perturbative QCD) vanishes completely.
Since no region of the phase space is forbidden, the cancellation of soft gluon effects takes place as in ordinary infrared-safe cross sections. 
That is the main advantage of this criterion: it eliminates all the fragmentation component in an infrared-safe way.
By contrast, it can not be implemented within the usual experimental conditions;
the standard way of implementing isolation in experiments is to use the prescription of Eq.~\eqref{Eq:Isol_chinormal} with a constant
$\chi(r)=1$. In any case, from a purely pragmatic point of view, it has been recently shown~\cite{Butterworth:2014efa} that
if the isolation parameters are tight enough (e.g., $E_{T~max} < 6~$GeV, $R=0.4$), the standard and the smooth cone isolation prescription
coincide at the $1\%$ level, which is well within the theoretical uncertainty of our predictions.

The acceptance criteria used in this analysis ($\sqrt{\rm s}=7$~TeV) are those implemented by the ATLAS collaboration analysis \cite{Aad:2012tba};
in  all the numerical results presented in this paper, we require $p_T^{\rm harder} \geq 25$~GeV, $p_T^{\rm softer}\geq 22$~GeV, and
we restrict the rapidity of both photons to satisfy $|y_\gamma|<1.37$ and \mbox{$1.52<|y_\gamma| \leq 2.37$}. 
The isolation parameters are set to the values $E_{T~max}=4~$GeV, $n=1$ and $R=0.4$, and the minimum angular separation between the two photons is $R_{\gg}=0.4$. We use the Martin-Stirling-Thorne-Watt (MSTW) 2008 \cite{Martin:2009iq} sets of parton distributions, with
densities and $\as$ evaluated at each corresponding order
(i.e., we use $(n+1)$-loop $\as$ at N$^n$LO, with $n=0,1,2$),
and we consider $N_f=5$ massless quarks/antiquarks and gluons in 
the initial state. The default
renormalization ($\mu_R$) and factorization ($\mu_F$) scales are set to the value
of the invariant mass of the diphoton system,
$\mu_R=\mu_F = M_{\gamma\gamma}$, while the default resummation scale ($\mu_{res}$) is set to $\mu_{res} = \Mgg/2$.
The QED coupling constant $\alpha$ is fixed to $\alpha=1/137$. 
\begin{figure}[htb!]
\begin{center}
\begin{tabular}{cc}
\includegraphics[width=0.47\textwidth]{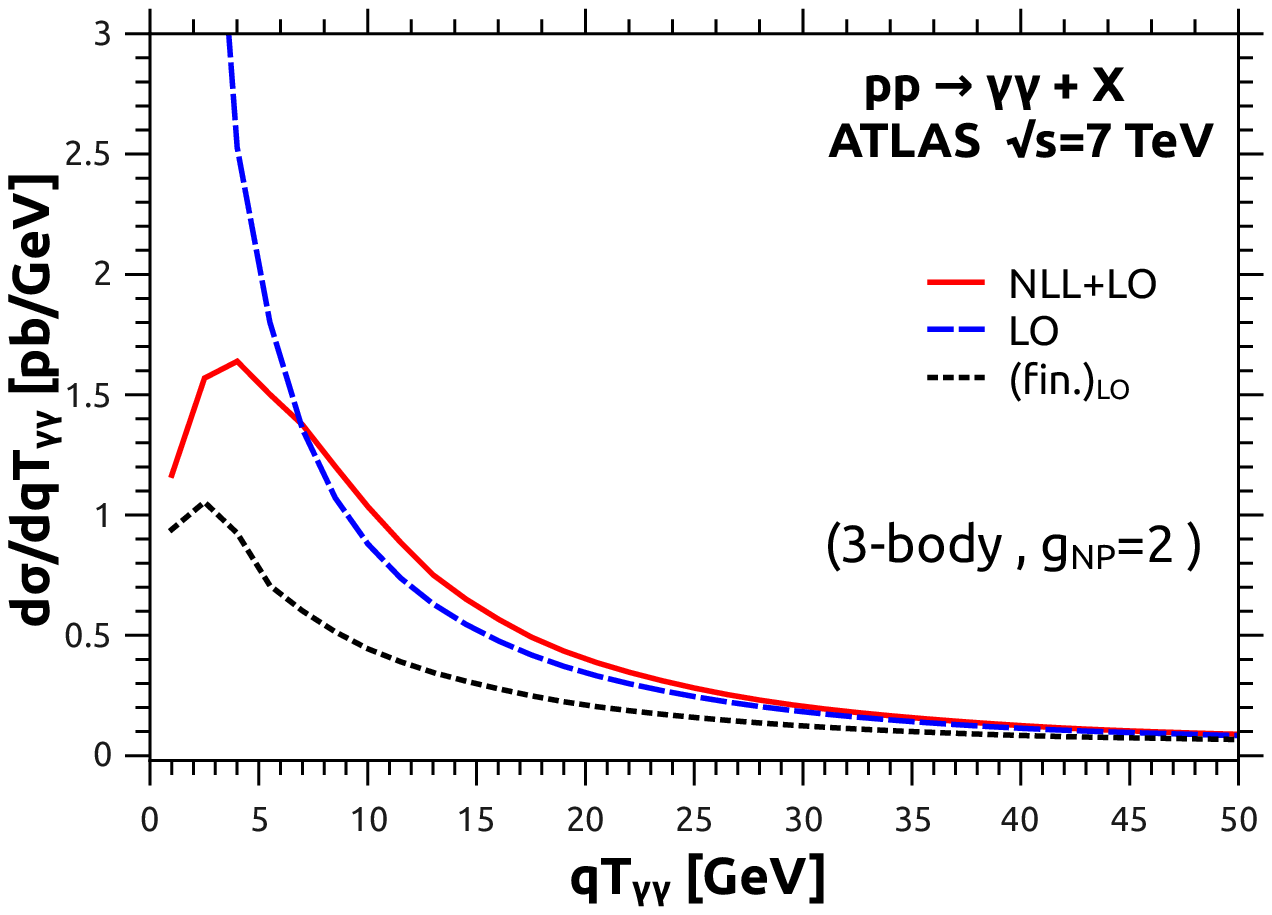} & \includegraphics[width=0.47\textwidth]{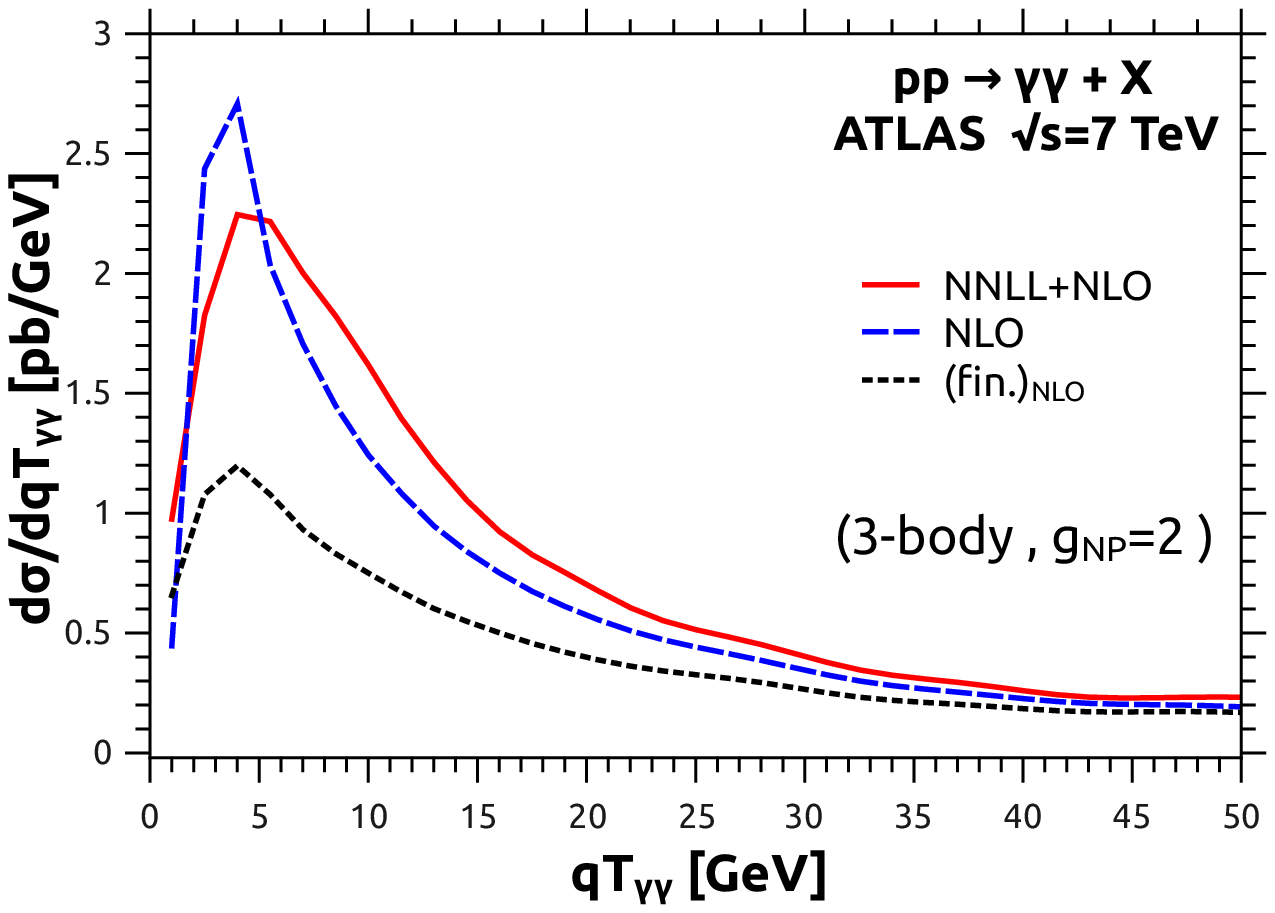}\\
\end{tabular}
\end{center}
\caption{\label{fig1}
{\em The $q_T$ spectrum of the photon pair (solid red lines) at the LHC (7~TeV): results at NLL+LO 
(left panel) and NNLL+NLO (right panel) accuracy. Each result is compared to the corresponding
fixed-order result (dashed lines) and to the finite component (dotted lines) in Eq.~(\ref{resfin}). 
The resummed
spectrum includes a non-perturbative (NP) contribution parametrized as in Eq.~\eqref{NPcon} and 
is obtained within the 3-body approach.}}
\end{figure}

This choice of the order of the parton densities and $\as$ is fully justified both in the small-$q_T$ region
(where the calculation of the partonic cross section includes the complete NNLO (NLO) result and is controlled by NNLL (NLL) 
resummation) and in the intermediate-$q_T$ region
(where the calculation is  constrained by the value of the NNLO (NLO) total cross section).

Non-perturbative (NP) effects are expected to be important at very small $q_T$. 
In this paper we follow the strategy of Ref.~\cite{Bozzi:2005wk}, implementing them
by multiplying  the $b$-space form
factor $\mathcal{W}^{\gg}$ of Eq.~\eqref{resum} by a `NP factor' which consists of a 
gaussian smearing of the form
\begin{equation}
\label{NPcon}
S^a_{NP} = \exp( -C_a \, g_{NP} \, b^2),
\end{equation}
where $a$ denotes the initial state channel, $a=F$ for $q\bar{q}$ and $a=A$ for $gg$ (as usual,
$C_F = (N_c^2 -1)/(2 N_c)$ and $C_A = N_c$).
In order to asses the importance of the NP contributions, we vary $g_{NP}$ in the interval
from $g_{NP} = 0$~GeV$^2$ (no NP contributions)
to $g_{NP} = 2$~GeV$^2$, corresponding to \textit{moderate} NP effects~\cite{Bozzi:2005wk}.

An additional and potentially important source of theoretical uncertainty arises from an 
ambiguity in the definition of the photon momenta
in the resummation formalism.
In fact, in the main resummation formula \eqref{resum}, which is used to define both the resummed
and (via the subtraction Eq.~\eqref{resfin})
the finite contributions to the partonic cross-section,
the diphoton pair total transverse momentum $q_T$ is not associated with the recoil of any extra 
physical particles in the final state. After $q_T$ resummation the angular distributions of the 
photons are still provided by the Born level functions ($\sigma_{\gg}^{(0)}$, ${\cal H}_{N_1 N_2}^{\gg}$),
which appear as multiplicative factors in front of the Sudakov form factor of Eq.~\eqref{exponent}. 
At this point there are two strategies to follow, which differ by corrections that are of $\mathcal{O}(q_T/\Mgg)$ order-by-order in the 
perturbative expansion~\cite{new:Daniel} (after having
matched the resummed calculation with a complete $N^kLO$ calculation, these corrections start to
contribute at the $N^{k+1}LO$).

\begin{figure}[htb]
\begin{center}
\begin{tabular}{cc}
\includegraphics[width=0.48\textwidth]{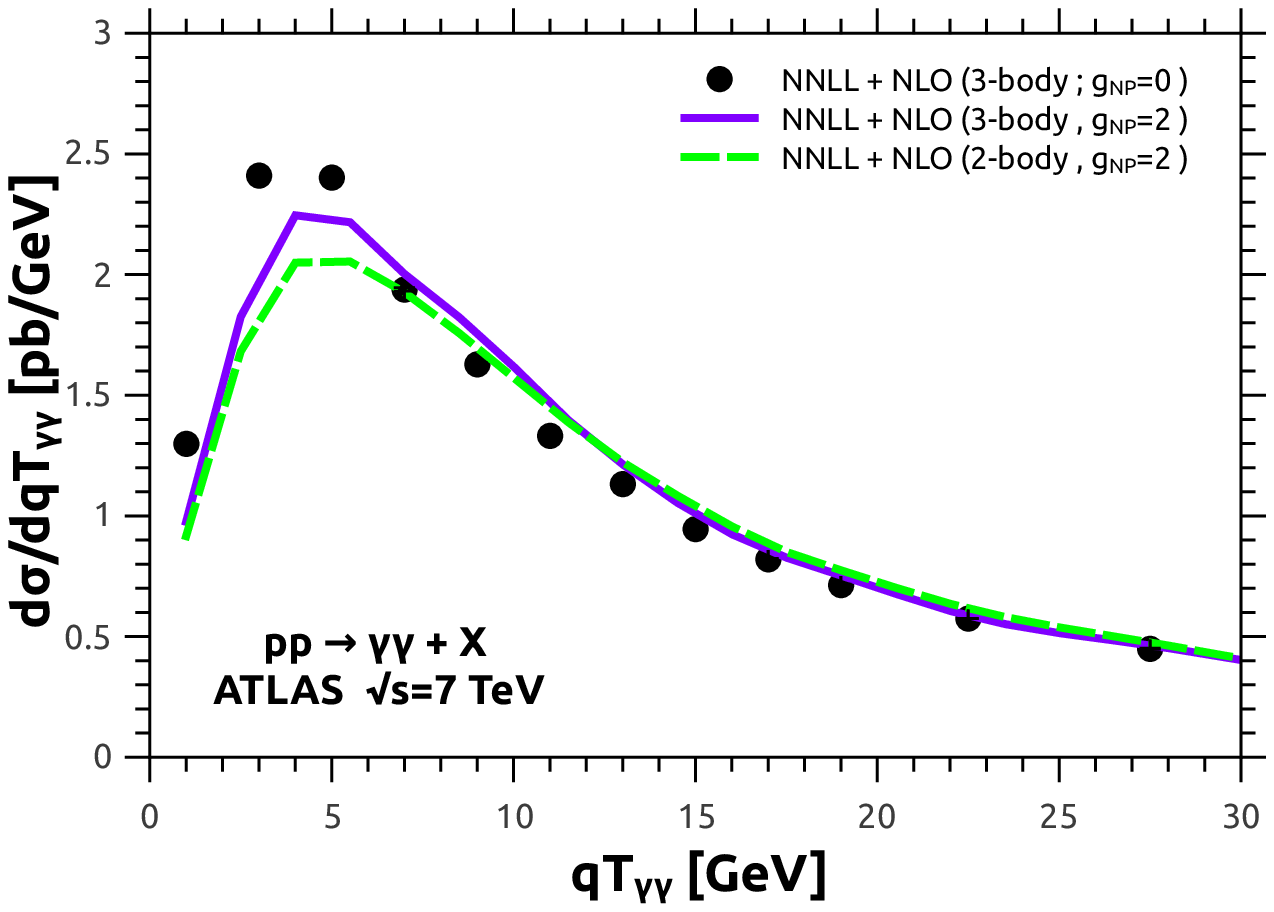} & \includegraphics[width=0.48\textwidth]{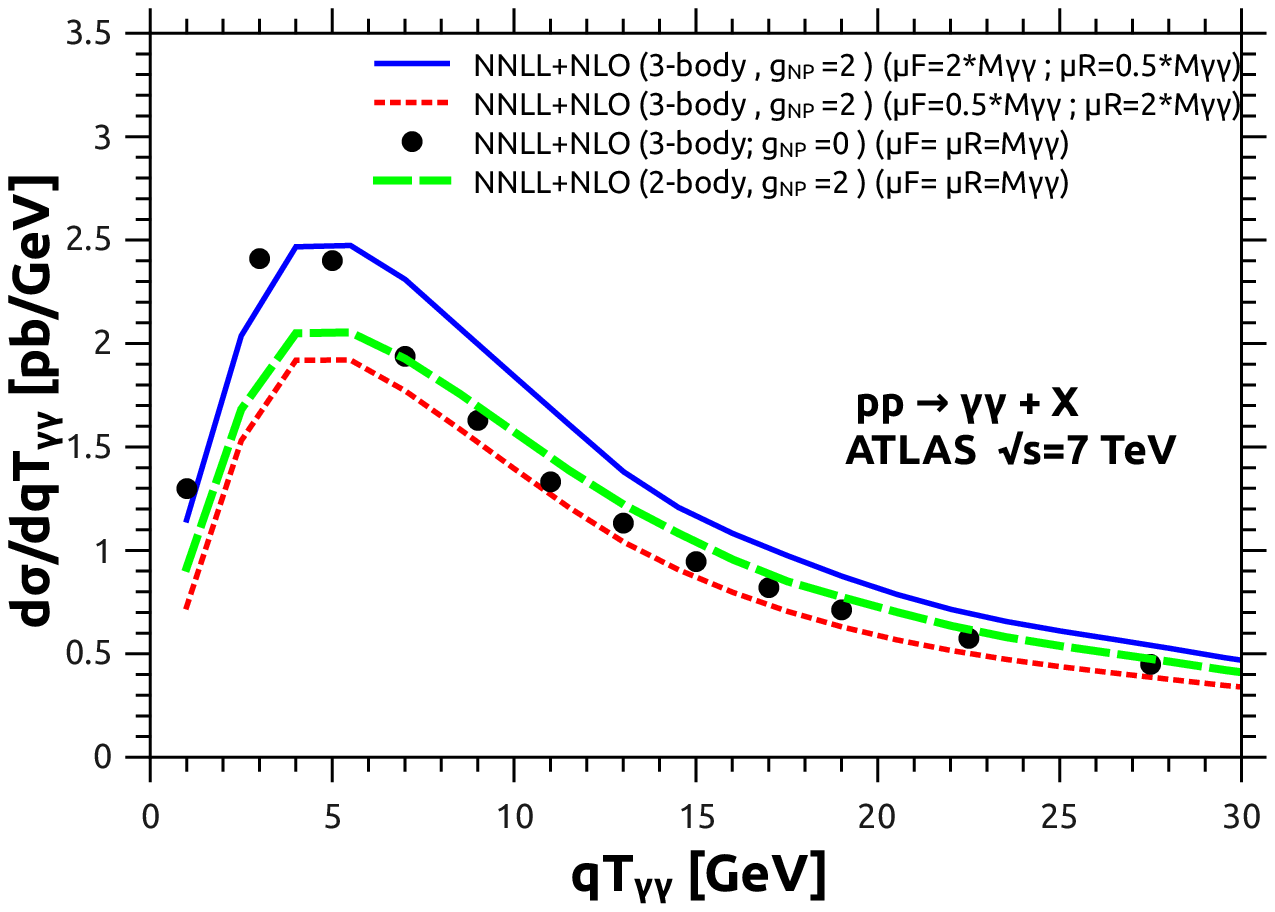}\\
\end{tabular}
\end{center}
\caption{\label{fig2}
{\em Uncertainties in the $q_T$ spectrum of diphoton production at the LHC ($\sqrt{s}=7$~TeV). In the left panel we show the different predictions associated with the 2-body and 3-body parametrization of the photon momenta at the central scale $\mu_F=\mu_R=\Mgg$, $\mu_{res}=\Mgg/2$. In the right panel,
the bands (solid and dotted lines) are obtained by varying $\mu_F$ and $\mu_R$ as described in the text; the prediction at the central scale with the NP contribution turned off ($g_{NP} = 0$~GeV$^2$ in Eq.~\eqref{NPcon}) is also shown.}}
\end{figure}
One of them is to just use the Born phase space for all the angular distributions in front of the Sudakov form factor of Eq.~\eqref{exponent}. In this case the transverse momentum $q_T$ is neglected in the calculation of the photon momenta, while the momenta $q_{i}\,,~(i=1,2)$ of the colliding partons are given by:
\begin{equation}
q_1^{\mu}=x_1 P_1^{\mu} \;\;\;\;\;\;\;\;\;\; q_2^{\mu}=x_2 P_2^{\mu}\;,
\label{2-body_LO}
\end{equation}
where $x_i\,~(i=1,2)$ are the parton momentum fractions and $P_i\,~(i=1,2)$ are the momenta of the colliding hadrons. The momenta $q_i$ respects the Born level kinematics
\begin{equation}
q_1 + q_2 = q_{\gamma 1} + q_{\gamma 2} \;\;,
\label{2-body}
\end{equation}
where  $q_{\gamma i}\,,~(i=1,2)$ are the 
momenta of the two photons in the final state. In this case the two photons are always in the
\textit{back-to-back} configuration, and therefore the kinematic effects of the transverse-momentum recoil, are not included in the momentum of each single photon. We call this approach the 2-body phase space; all the differential distributions which use this method have a (2-body) label.

A more elaborate, and arguably more physical, approach is to consider the effects of the transverse-momentum recoil in the two-photon final state. Therefore these kinematic effects have to be `absorbed' by the incoming parton momenta $q_1$ and $q_2$, in order to respect the Born level kinematics
\begin{equation}
q_1(q_T) + q_2(q_T) = q_{\gamma 1}(q_T) + q_{\gamma 2}(q_T) \;\;,
\label{3_body_qt}
\end{equation}
and the LO kinematics of Eq.~\eqref{2-body_LO}. There are different consistent implementations of this approach, all of them giving equivalent (up to higher order corrections) results~\footnote{The details of the implementation of these kinematic effects are discussed in a forthcoming paper~\cite{new:Daniel}.}. The implementation that we use in the following is essentially equivalent to define the separate photon momenta according to the Born level angular distribution computed in the Collins-Soper frame~\cite{Collins:1977iv} of the diphoton system (in practice, this is the same procedure used in Ref.~\cite{Balazs:2007hr}).
We call this method the 3-body phase space, and distributions obtained using this method are labelled (3-body). The modifications introduced by this phase space parametrization in the momentum fractions $x_1$, $x_2$ of the incoming partons are neglected because they produce negligible effects ($\mathcal{O}(q_T/\Mgg)$) in the cross-section.
Notice that it is not the formal 3-body phase space (the 3-body effect is produced by the Lorentz transformation with finite $q_T$ of the Collins-Soper frame).
We have checked that if the formal 3-body phase space is used~\footnote{Which violates the Born level kinematics of Eq.~\eqref{2-body}.}, the changes in the cross section (comparing the 2-body and 3-body cases) are of order $\mathcal{O}(q_T/\Mgg)$.
The ambiguity in the treatment of the photon momenta is considered as an additional source of theoretical uncertainty.
\begin{figure}[htb]
\begin{center}
\begin{tabular}{cc}
\includegraphics[width=0.48\textwidth]{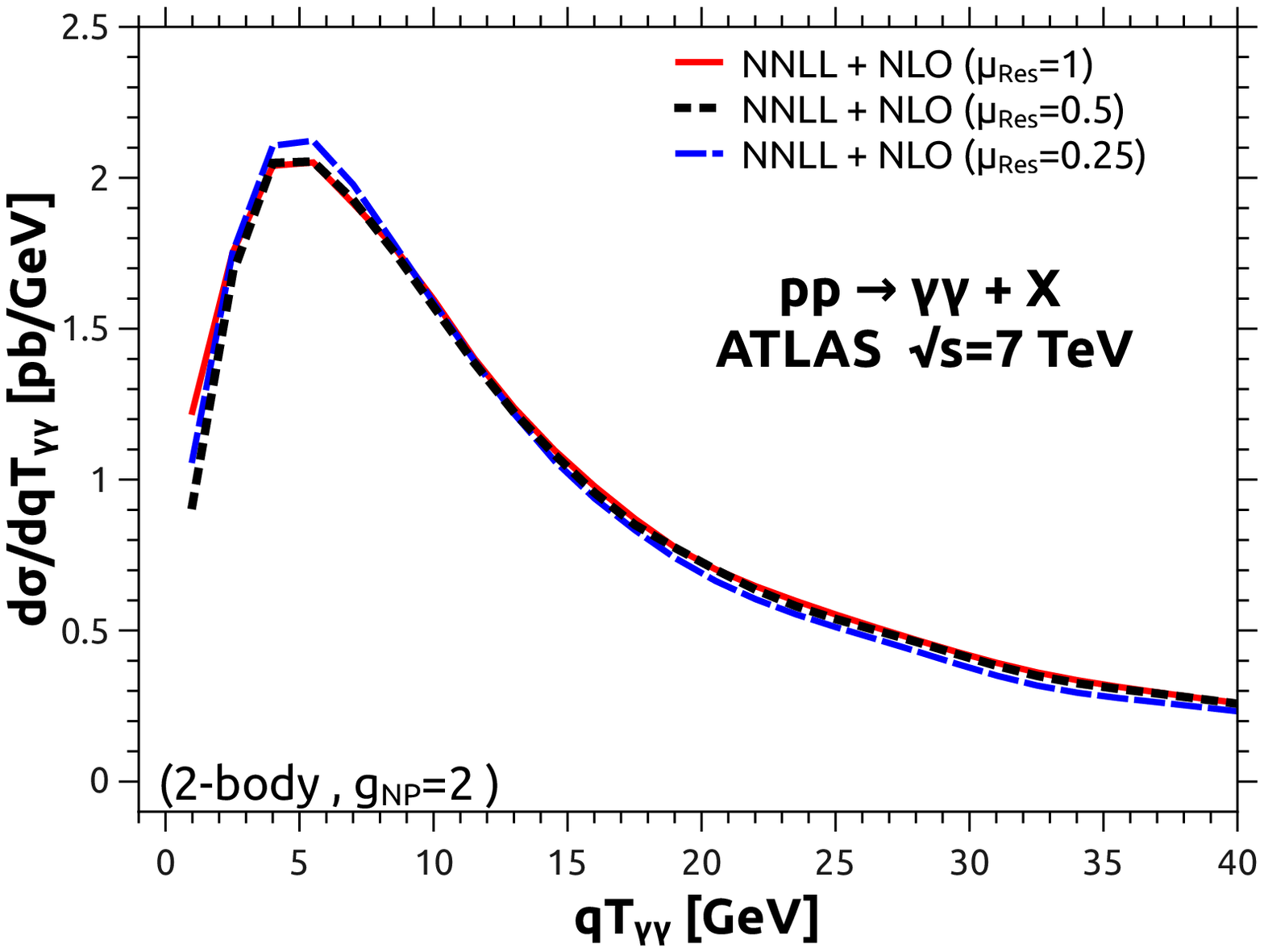} & \includegraphics[width=0.48\textwidth]{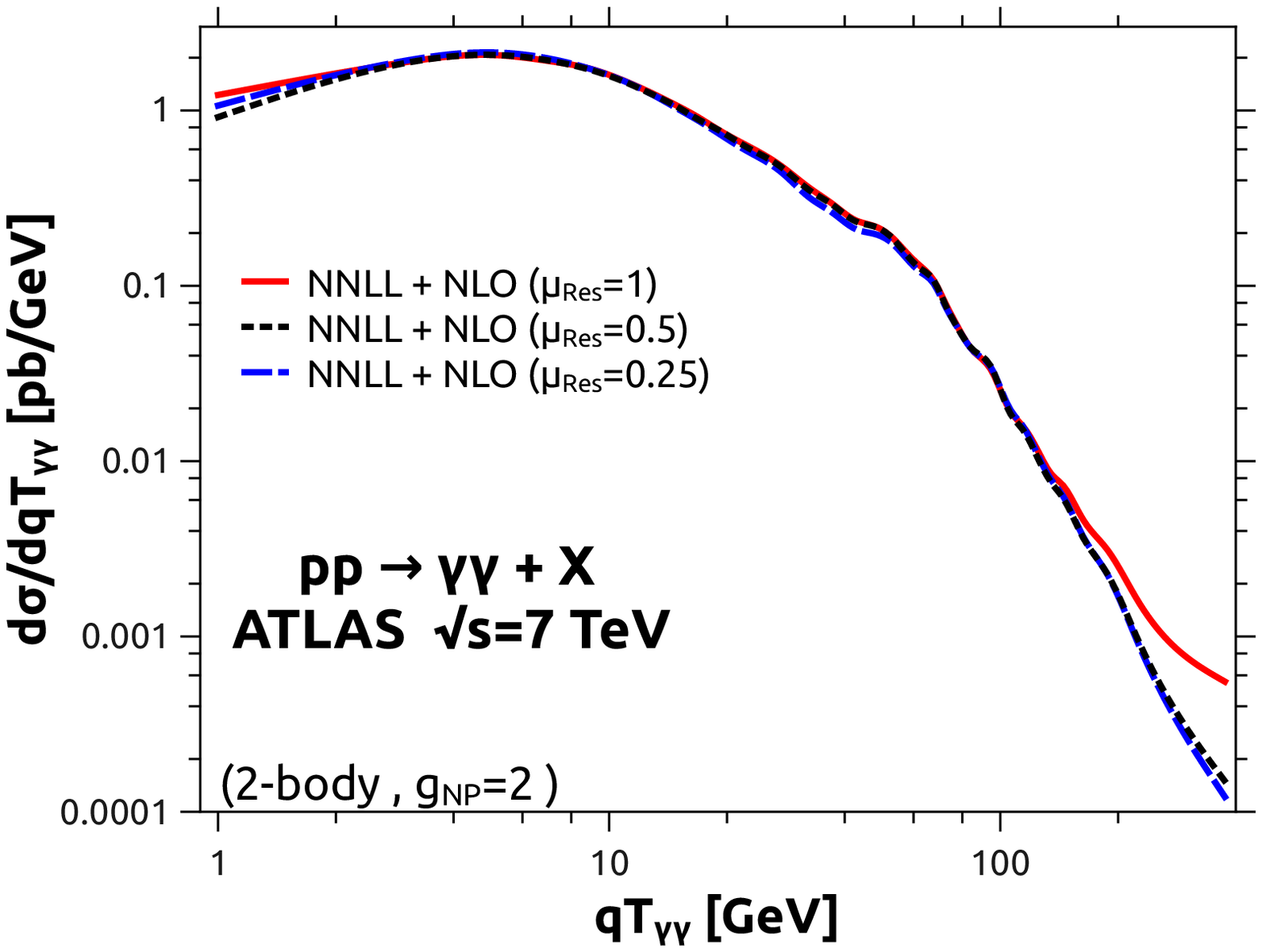}\\
\end{tabular}
\end{center}
\caption{\label{fig2_2}
{\em The $q_T$ spectrum of diphoton production at the LHC ($\sqrt{s}=7$~TeV). Here, the scales 
 $\mu_F=\mu_R=\Mgg$ are kept fixed while we vary the resummation scale $\mu_{res}$ 
 to establish its impact on the cross section. In the left panel we show the 
 range $0$~GeV$<q_{T\gg}<40$~GeV, and in the right panel the full spectra in logarithmic scale.}}
\end{figure}

In Fig.~\ref{fig1}, left panel, we present  the NLL+LO $q_T$ spectrum at the LHC ($\sqrt{s}=7$~TeV).
The NLL+LO result (solid line) at the default scales ($\mu_F=\mu_R=\Mgg$;~$\mu_{res}=\Mgg/2$) is 
compared with the
corresponding LO result (dashed line). We use the 3-body phase space and a NP parameter $g_{NP}=2$~GeV$^2$.
The LO finite component of the spectrum (see Eq.~(\ref{resplusfin})), is also shown 
for comparison (dotted line).
We observe that the LO result diverges to $+\infty$ as $q_T\to 0$, as expected. The finite component is regular over the full $q_T$ range, it smoothly vanishes as $q_T\to 0$ and gives an important contribution to the NLL+LO result in the low-$q_T$ region.
That is mostly originated by the $qg$ channel, which starts at NLO and provides a {\it subleading} 
correction in terms of logs (single logarithmic terms) but contributes considerably to the cross-section
due to the huge partonic luminosity compared to the formally leading $q\bar{q}$ channel.
The resummation of the small-$q_T$ logarithms
leads to a well-behaved distribution: it vanishes as $q_T\to 0$ and approaches the corresponding LO result
at large values of $q_T$.

The results in the right panel of Fig.~\ref{fig1} are systematically at one order
higher: the $q_T$ spectrum at NNLL+NLO accuracy (solid line) is compared with
the NLO result (dashed line) and with the NLO finite component of the spectrum
(dotted line).
The NLO result diverges to $-\infty$ as $q_T\to 0$ and, at small values of $q_T$,
it has an unphysical peak that is produced by the compensation of negative leading
and positive subleading logarithmic contributions.
The contribution of the NLO finite component to the NNLL+NLO result is of the order of the 50\% at the peak and becomes more important as $q_T$ increases.
A similar quantitative behaviour is observed by considering the contribution of
the NLO finite component to the NLO result.
At large values of $q_T$ the contribution of the NLO finite component tends 
to the NLO result. This behaviour indicates that the logarithmic terms are no 
longer dominant and that the resummed
calculation cannot improve upon the predictivity of the fixed-order expansion. 

We also observe that the position of the peak in
the NNLL+NLO $q_T$ distribution is slightly harder than the corresponding 
NLL+LO $q_T$ distribution. This effect is (in part) due to the large 
transverse-momentum dependence of the fixed order corrections.

As discussed in Sect.~\ref{sec:theory}, the resummed calculation depends on the factorization and 
renormalization scales and on the resummation scale $\mures$. 
Our convention to compute factorization  and renormalization scale uncertainties is to consider 
independent variations of $\mu_F$ and $\mu_R$ by a factor of two around 
the central values $\mu_F=\mu_R=\Mgg$ in independent way in order to maximise them: ($\mu_F=2~\Mgg$, $\mu_R=\Mgg/2, \mures=\Mgg/2$)
and ($\mu_R=2~\Mgg$, $\mu_F=\Mgg/2, \mures=\Mgg/2$). The uncertainty due to the resummation scale variation
is assessed separately by varying it between $\mu_{res} = \Mgg/4$ and $\mu_{res} = \Mgg$ at fixed $\mu_F$ and $\mu_R$.

\begin{figure}[htb]
\begin{center}
\begin{tabular}{cc}
\includegraphics[width=0.47\textwidth]{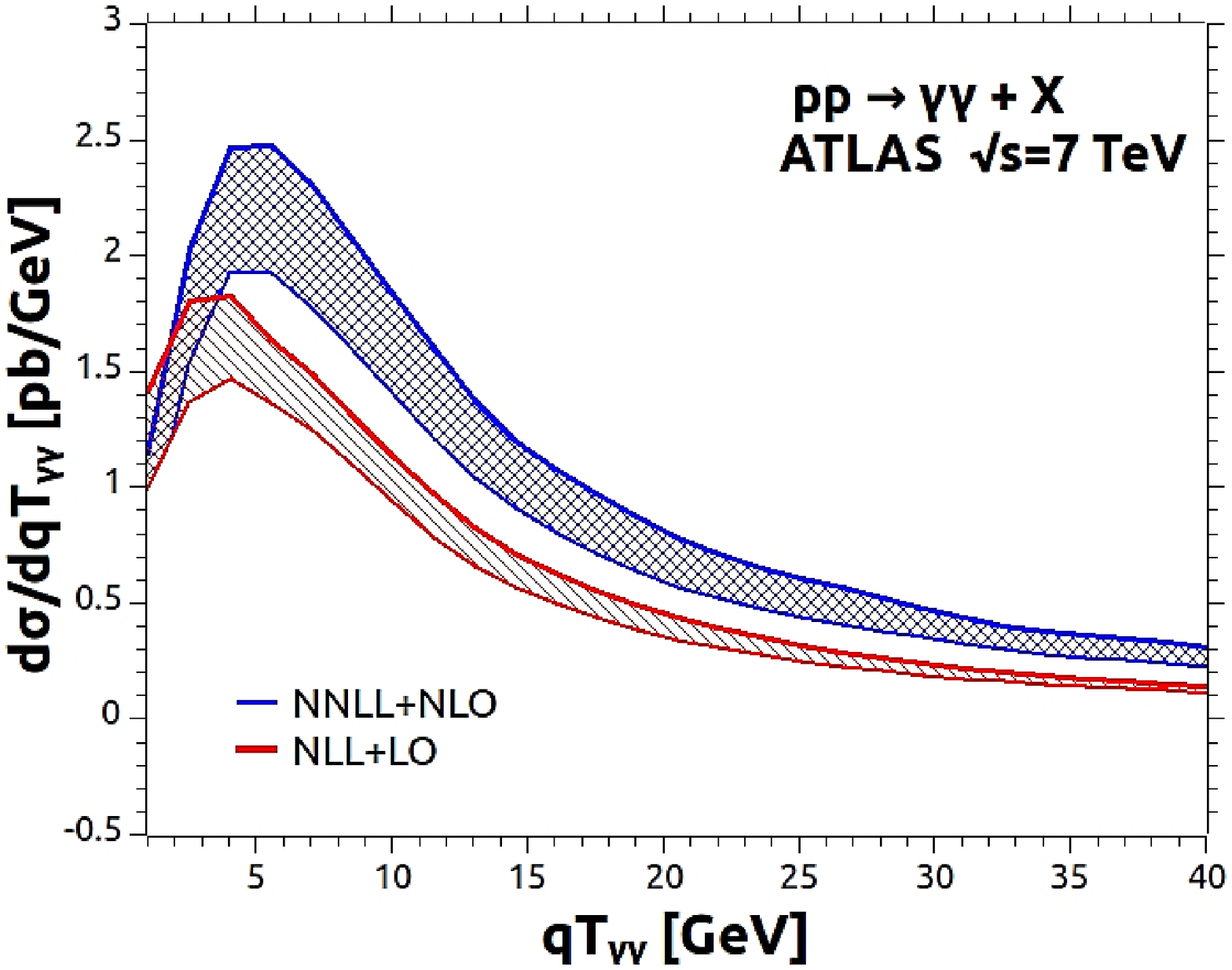} & \includegraphics[width=0.47\textwidth]{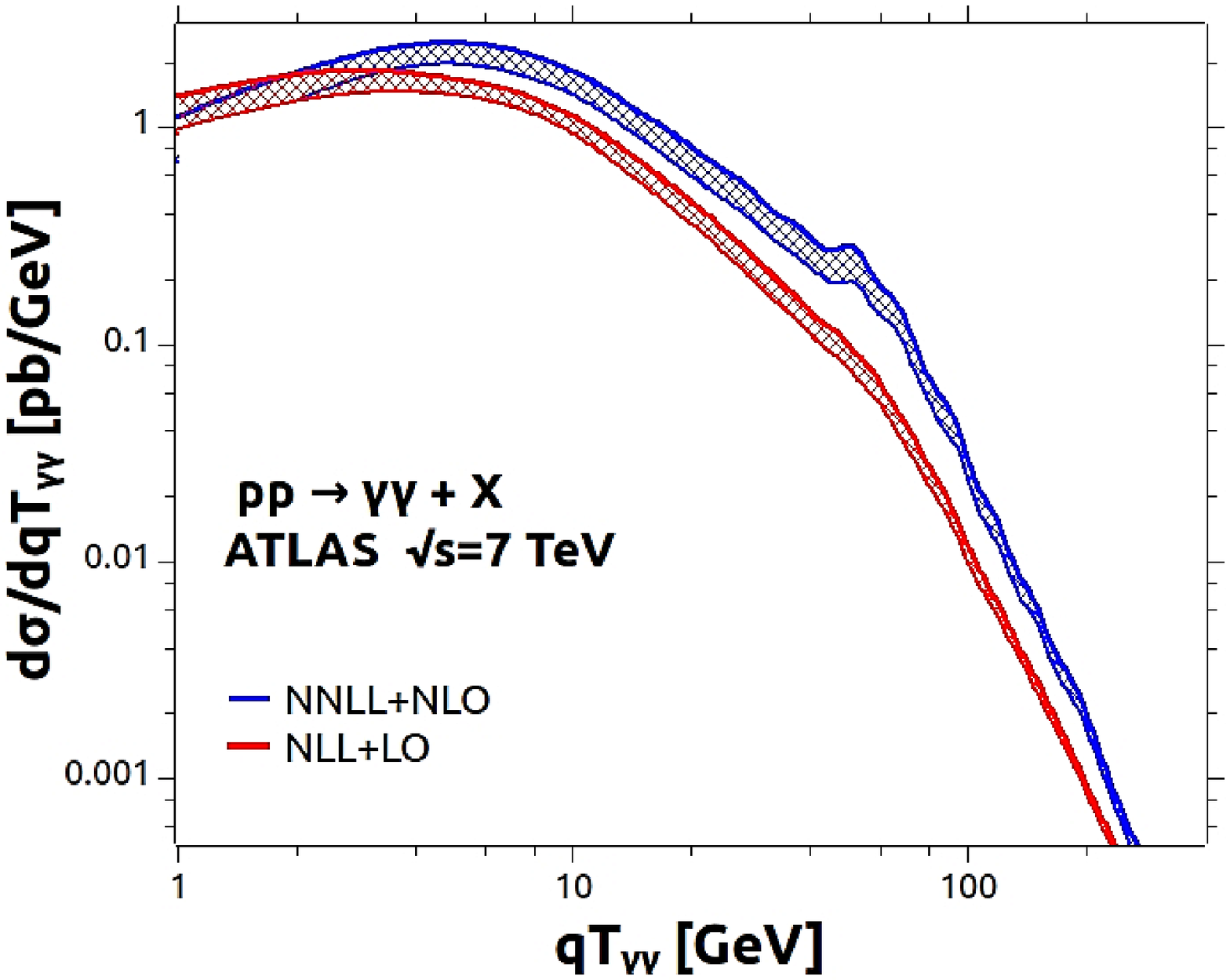}\\
\end{tabular}
\end{center}
\caption{\label{fig3}
{\em The $q_T$ spectrum of diphoton pairs at the LHC. The
NNLL+NLO result is compared with the NLL+LO result, for the window 0 GeV $< q_T<40$~GeV (left panel)
and the full spectra (right panel). We use the
3-body parametrization in the resummed cross-section, and set $g_{NP} = 2$~GeV$^2$. The bands are obtained
by varying $\mu_R$ and $\mu_F$ as explained
in the text.}}
\end{figure}

\begin{figure}[htb]
\begin{center}
\begin{tabular}{cc}
\includegraphics[width=0.48\textwidth]{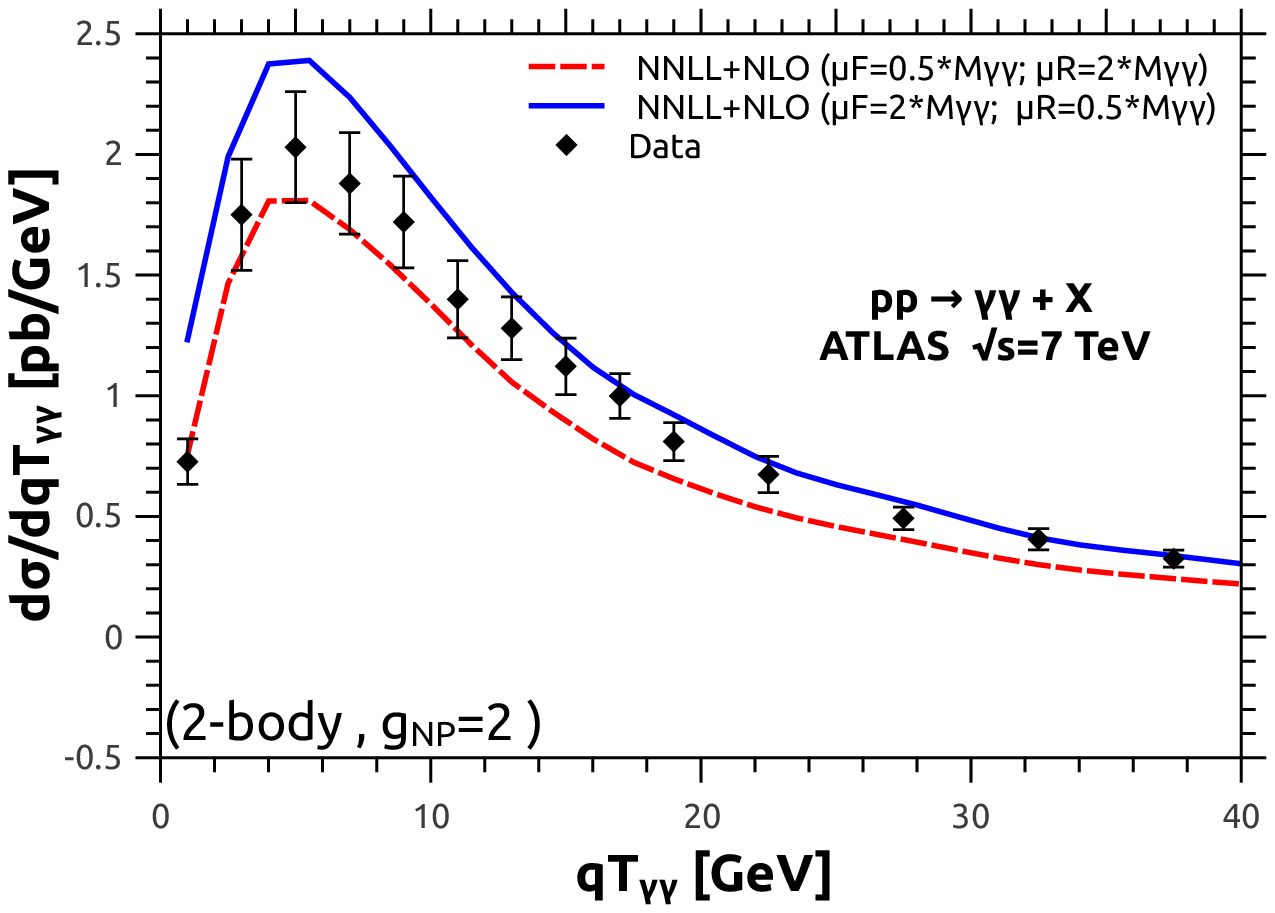} & \includegraphics[width=0.48\textwidth]{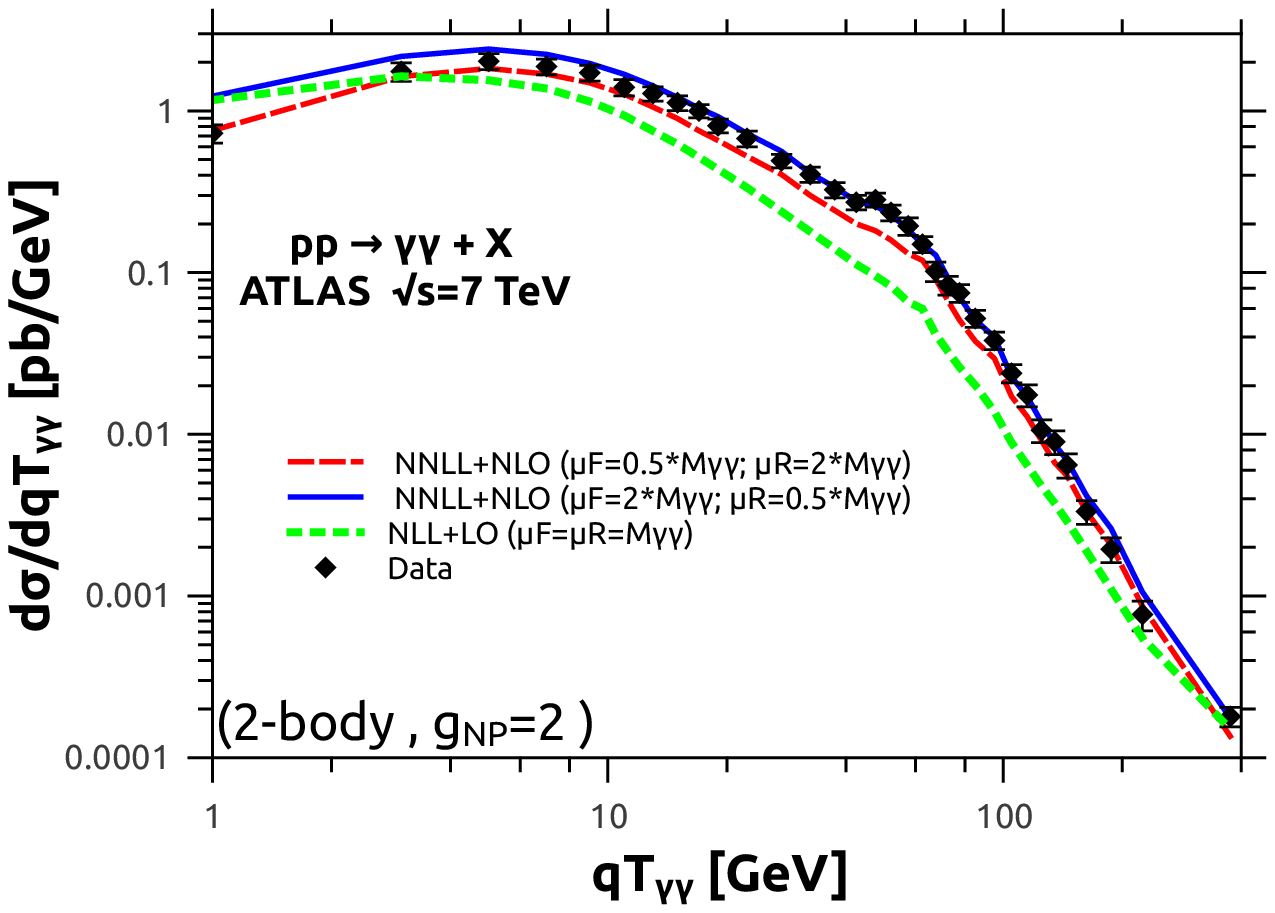}\\
\end{tabular}
\end{center}
\caption{\label{fig4}
{\em Comparison of the theoretical prediction for the $q_T$ spectrum of diphoton pairs at the LHC to the experimental data. The
NNLL+NLO result is compared with the ATLAS data of Ref.~\cite{Aad:2012tba}, for the window 0 GeV $< q_T<40$~GeV (left panel)
and the full spectra (right panel). In the right panel the NNL+LO distribution at central scale (dotted line) is also shown in order to compare it with the data and the NNLL+NLO result.
In the theoretical curves we use the 2-body parametrization
for the resummed cross-section and set $g_{NP} = 2$~GeV$^2$; the bands (solid and dashed lines) are obtained by varying $\mu_R$ and $\mu_F$ as explained
in the text.}}
\end{figure}

\begin{figure}[htb]
\begin{center}
\begin{tabular}{cc}
\includegraphics[width=0.48\textwidth]{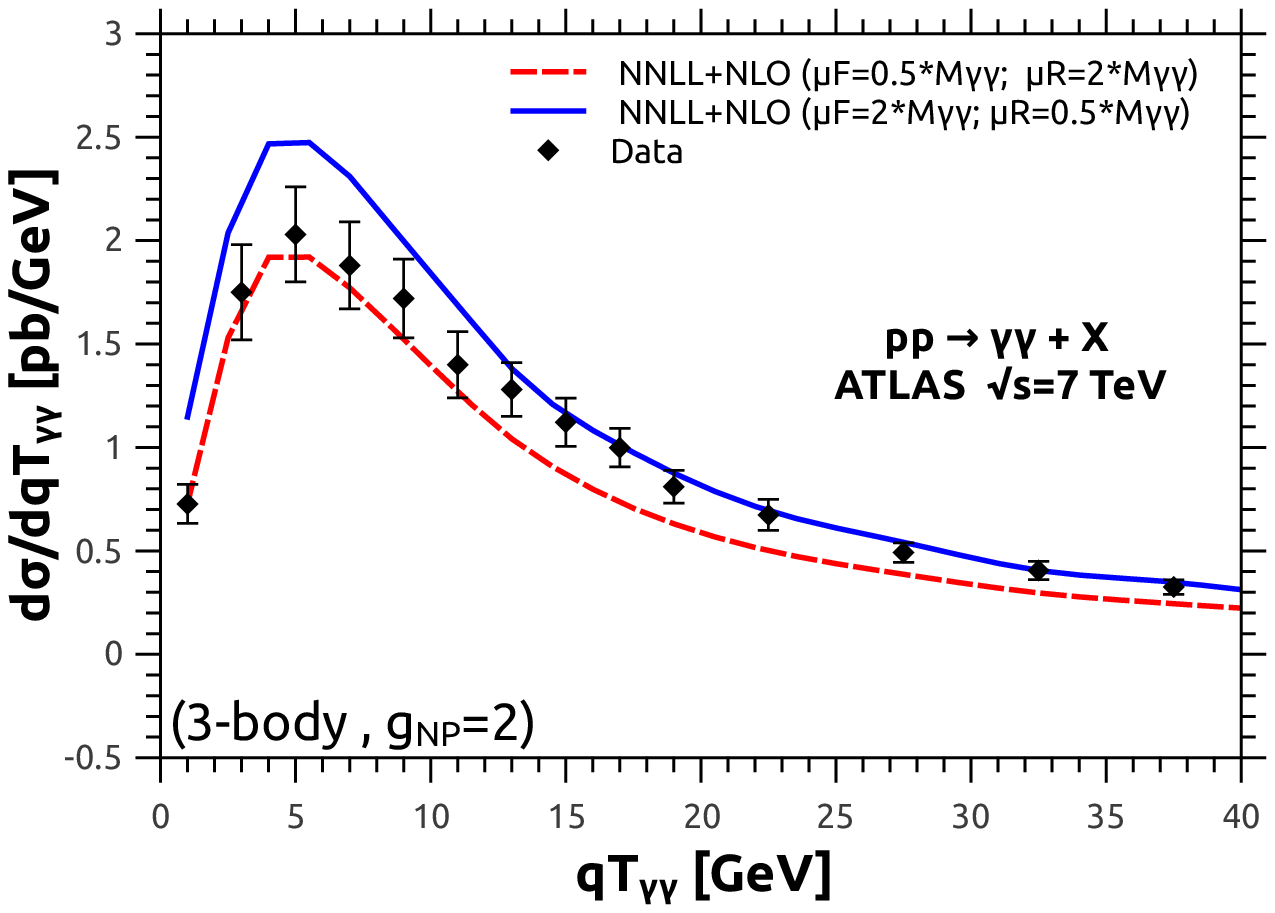} & \includegraphics[width=0.48\textwidth]{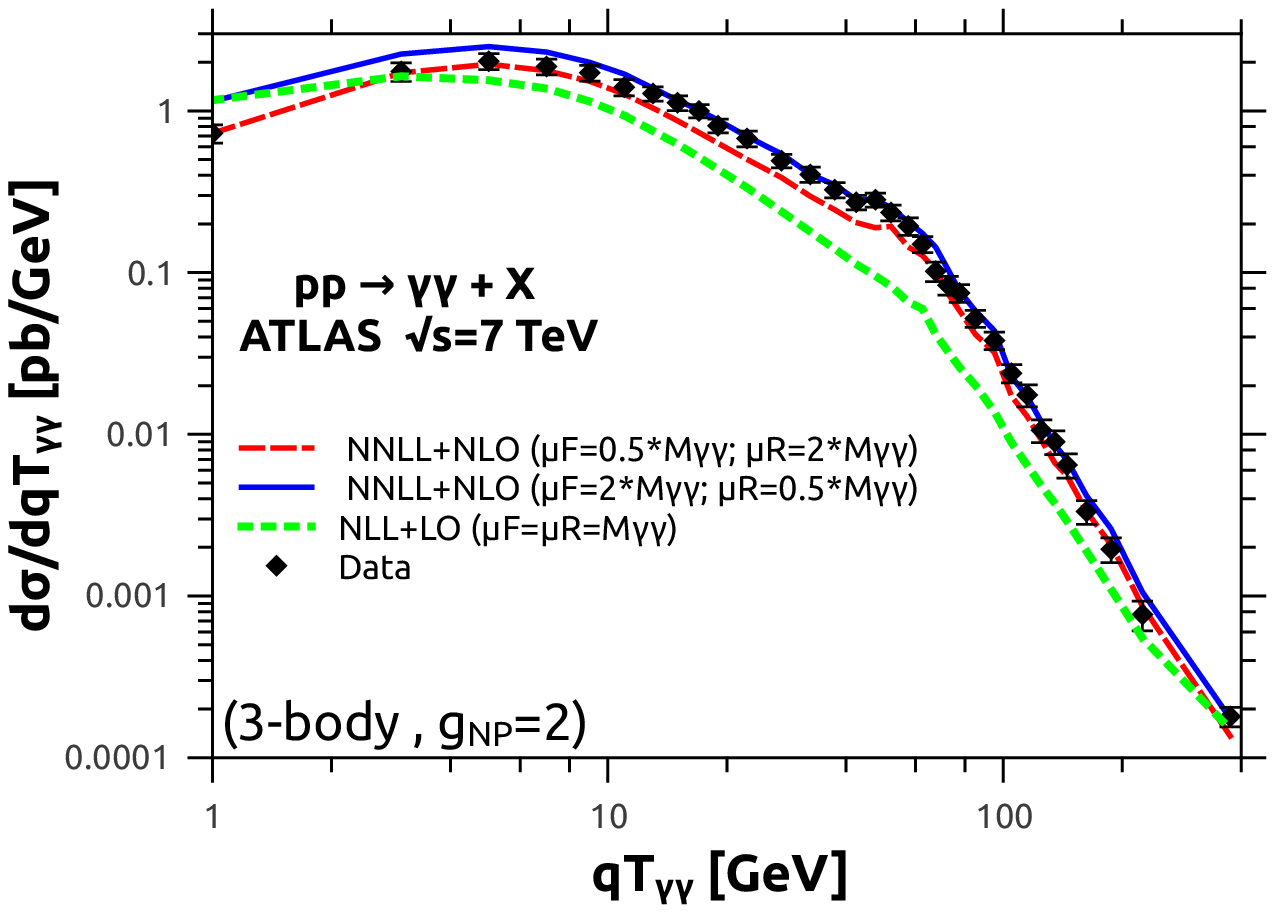}\\
\end{tabular}
\end{center}
\caption{\label{fig5}
{\em Comparison of the theoretical prediction for the $q_T$ spectrum of diphoton pairs at the LHC with the experimental data. The
NNLL+NLO result is compared with the ATLAS data of Ref.~\cite{Aad:2012tba}, for the window 0 GeV $< q_T<40$~GeV (left panel) and
the full spectra (right panel). In the right panel the NNL+LO distribution at central scale (dotted line) is also shown in order to compare it with the data and the NNLL+NLO result.
In the theoretical curves we use the 3-body parametrization
for the resummed cross-section and set $g_{NP} = 2$~GeV$^2$. The bands (solid and dashed lines) are obtained by varying $\mu_R$ and $\mu_F$ as explained
in the text.}}
\end{figure}


\begin{figure}[htb]
\begin{center}
\begin{tabular}{cc}
\includegraphics[width=0.48\textwidth]{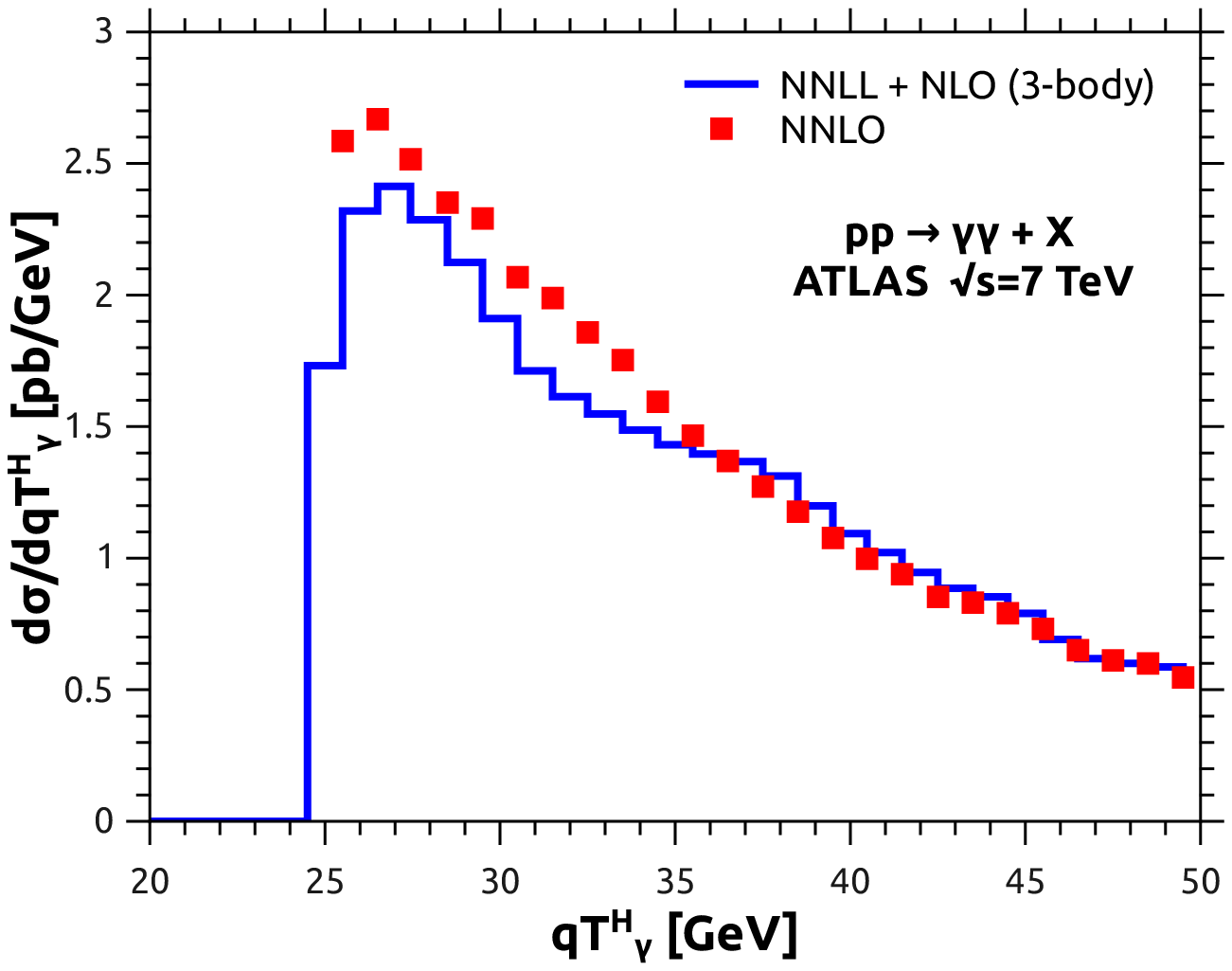} & \includegraphics[width=0.48\textwidth]{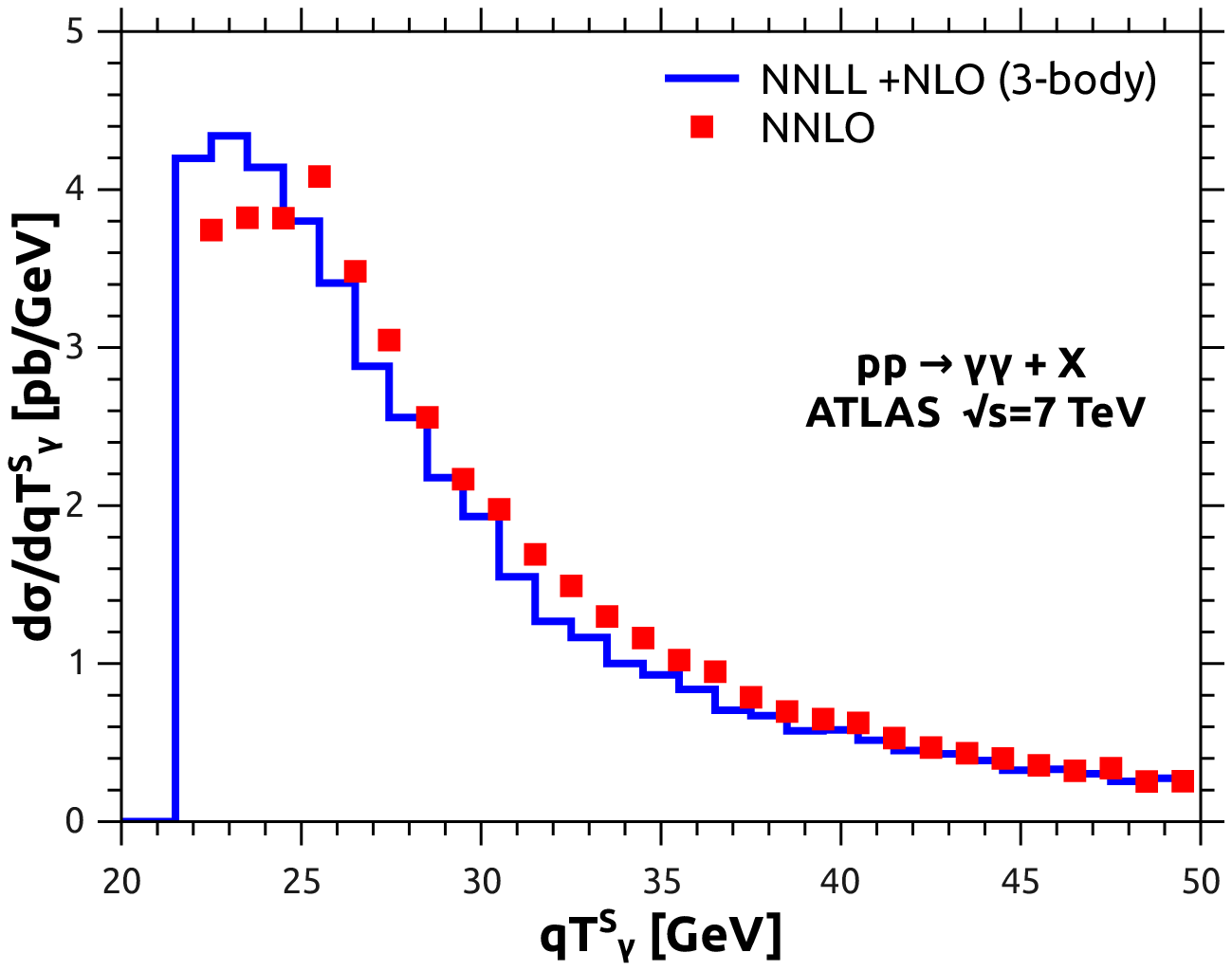}\\
\end{tabular}
\end{center}
\caption{\label{fig5_1}
{\em Individual photon spectra in diphoton production at the LHC. Left panel: the $q_{T\gamma}^{H}$ spectrum of the hard photon at the LHC~\cite{Aad:2012tba} ($\sqrt{s}=7$~TeV). Right panel: the $q_{T\gamma}^{S}$ spectrum of the softer photon at the LHC~\cite{Aad:2012tba} ($\sqrt{s}=7$~TeV). In
both panels, we compare the resummed prediction using the 3-body parametrization in the resummed cross-section ($g_{NP}=2$~GeV$^2$) with the fixed order prediction.}}
\end{figure}
In Fig.~\ref{fig2} we compare the impact of the various sources of theoretical error for the NNLL+NLO
predictions: the 2-body and 3-body
parametrizations of the photon phase space, the effect of the variation of the non-perturbative 
contribution, and the variation of the factorization
and renormalization scales.
In the left panel of Fig.~\ref{fig2} we use only the central scale ($\mu_F=\mu_R=\Mgg$, $\mures=\Mgg/2$), and we note 
that the
3-body method results in a slightly larger cross section (about 10\%) around the peak ($q_T\sim 5$~GeV) 
than the 2-body one. For $q_T> 20$~GeV all
the contributions in the left panel coincide, which is consistent with the fact that at these values
the resummed component starts to vanishes and the
fixed order result dominates the cross section, as we can anticipate from Fig.~\ref{fig1}. Also we note 
that if we use a NP parameter, the peak of
the distribution is located at larger values of $q_T$ ($q_T\sim 5$~GeV) and the shape of the
distribution is slightly different (for values 0~GeV$<q_T < 22$~GeV)
from the case in which the NP parameter is not implemented. These differences, which are stronger for $q_T < 10$~GeV, have their origin in the resummed component, which is the only contribution that depends on $g_{NP}$. 
In the right panel of Fig.~\ref{fig2} 
we show the comparison between the
variation of the scales\footnote{We vary the scales from ($\mu_F=2\Mgg;\mu_R=\Mgg/2$) to 
($\mu_R=2\Mgg;\mu_F=\Mgg/2$) at fixed $\mures=\Mgg/2$.} in the 3-body approach
with the central scale results of the 2-body and 3-body frameworks (with and without non perturbative
parameter, respectively). Evidently,  the uncertainty
due to the ambiguity of the parametrization of the photon momenta turns out to be subdominant with respect 
to the one arising from the variation of the scales which provides, by itself, a reasonable estimate of 
theoretical uncertainties. 

In Fig.~\ref{fig2_2} we show the NNLL+NLO transverse momentum distribution for three different 
implementations of the $\mures$ parameter ($\mures=\Mgg/4;\Mgg/2;\Mgg$) at fixed $\mu_F=\mu_R=\Mgg$. The 2-body phase space and a non-perturbative parameter $g_{NP}=2$~GeV$^2$ were used. We notice the small impact of the variation of the $\mures$ scale in the cross section (at per-cent level). In the left panel of Fig.~\ref{fig2_2} we present the transverse momentum distribution for values of $q_T$ within the interval $0$~GeV$<q_T < 40$~GeV, and in the right panel of Fig.~\ref{fig2_2} the full spectra. 
We also notice that the strongest effect of the variation of the $\mures$ scale appears in the
last bin of right panel of Fig.~\ref{fig2_2}. This is expected since the resummation scale effectively {\it sets} the value of transverse momentum at which the logarithms are dominant. A choice of a very large resummation scale affects the distribution at larger transverse momentum and might in general result in a mismatch with the fixed order prediction due to the artificial introduction of unphysically large logarithmic contributions in that region. Similar results are obtained if the 3-body phase space is used instead of the 2-body one.

In Fig.~\ref{fig3} we compare the variation of the scales of the NNLL+NLO and NLL+LO predictions 
(3-body phase space), for the interval $0$~GeV$<q_T < 40$~GeV (left panel) and the full spectra (right panel). We notice that the dependence on the scales is not reduced when going 
from NLL+LO to NNLL+NLO. This is mostly because at NNLL+NLO a new channel (gg) opens,  in which 
the box contribution (effectively ``LO'' but formally $\mathcal{O}(\alpha_S^2)$) ruins the reduction of the scale dependence usually expected when adding second order corrections for the $q\bar{q}$ channel and first order corrections for the $qg$ channel. This effect has the same origin that the reported behaviour of the diphoton production at NNLO of Ref.~\cite{Catani:2011qz}, when the variation of the scales is implemented.
Since NNLL+NLO is the first order at which all partonic channels contribute, it is possible to argue that this is the first order at which estimates of theoretical uncertainties through scale variations can be considered as reliable. The same results are obtained if the 2-body phase space is used instead of the 3-body one.

The peak observed in the right panel of Fig.~\ref{fig3} is the so called
\textit{Guillet shoulder}~\cite{Binoth:2000zt}, which is a real radiation 
effect and has its origin in the fixed order contribution. It appears stronger in the 
NNLL+NLO $q_T$ distribution than in the NLL+LO, due to the larger size of the real contributions
at NLO.

In the small $q_T$ region ($q_T < 4$~GeV) the real radiation effects are no 
longer dominant in the $q_T$ distribution. The finite and resummed component
(which vanish as $q_T \rightarrow 0$) are of the same order for $q_T < 4$~GeV. 
In absence of the strong real radiation effects 
the contributions are almost completely Born like. This is the main reason why the NNLL+NLO 
bands overlaps with those at the previous order.

In Figs.~\ref{fig4} and \ref{fig5} we compare the LHC data ($\sqrt{s}=7$~TeV) from ATLAS~\cite{Aad:2012tba} with our resummed theoretical predictions (at NNLL+NLO and NLL+LO) using the 2-body and 3-body approaches, respectively.
In both cases we use a NP parameter different from zero ($g_{NP}=2$~GeV$^2$) and 
we estimate the theoretical uncertainty by the variation of the $\mu_R$ and $\mu_F$ scales.
In the left panels we show the $q_T$ distribution in the window (0~GeV$<q_T < 40$~GeV), 
while in the right ones we show the full spectra in logarithmic scale. 

We observe in general an excellent agreement between the resummed NNLL+NLO prediction and 
the experimental data, that is accurately described within the theoretical uncertainty bands 
in the whole kinematic range. Also we observe that the NLL+LO result is not enough to describe 
the phenomenology of the transverse-momentum distribution of the LHC data (Figs.~\ref{fig4} 
and \ref{fig5}, right panel). By direct comparison to the fixed order prediction, we notice
that the effect of resummation is not only to recover the predictivity of the calculation at
small transverse momentum, but also to improve substantially the agreement with 
LHC data~\cite{Aad:2012tba}.

While the resummation performed in this work reaches NNLL accuracy formally only for the diphoton transverse momentum distribution, its predictions can be extended to other observables as well, since at least the leading logarithmic contributions have a common origin from soft and collinear emission. Note that the 3-body approach is more suitable for a consistent implementation of these leading logarithmic effects on the observables.
In Fig.~\ref{fig5_1} we show results on more exclusive observables: the $q_T$ distributions of the 
harder (left-hand plot) and softer (right-hand plot) photon at the central scale. We compare the results obtained with the 2-body and 3-body phase space approaches, using a non-perturbative parameter $g_{NP}=2$~GeV$^2$ with the fixed order result at NNLO for diphoton production~\cite{Catani:2011qz}. 

The 2-body phase space transverse-momentum distribution at NNLL+NLO provides the same result than the NNLO fixed order cross section for diphoton production~\cite{Catani:2011qz}. This is consistent with two following related facts: \textit{i)} the single photon momentum does not carry any information about the recoil due to the transverse momentum $q_T$ in the 2-body approach; \textit{ii)} the NNLO (NLO) result for the total cross section is exactly recovered upon integration over $q_T$ of the differential cross section $d \sigma_{\gg}/dq_T$ at NNLL+NLO (NLL+LO) accuracy. 

The effects of resummation are only present in these more exclusive observables if the recoil due 
to $q_T$ are absorbed by the photons in the final state~\footnote{And also the recoil due to $q_T$ 
has to be absorbed in the initial state to restore the momentum conservation (see Eq.~\eqref{3_body_qt}).}, 
which is equivalent to the implementation of a 3-body like phase space. In this way, because the 
single photon momentum depends on $q_T$, the integral over $q_T$ of the NNLL+NLO (NLL+LO) distribution
does not recover the NNLO (NLO) result as we can observe in Fig.~\ref{fig5_1}. Here the last integral over the 
single photon momentum of the NNLL+NLO (NLL+LO) distribution, is required in order to recover the
NNLO (NLO) result for the total cross section.

We comment on the $q_T$ distribution of the softer photon in the region around the back-to-back threshold $q_{T\gamma}^{S} \sim 25$~GeV (see Fig.~\ref{fig5_1} right panel). The NLO fixed order result has a step-like behaviour, and this necessarily produces \cite{Catani:1997xc}
integrable logarithmic singularities at each subsequent perturbative order.
The peak of the NLO fixed order distribution at $q_{T~\gamma}^{S} \sim 25$~GeV
is an artifact of these perturbative instabilities. The instability is cured by
all-order perturbative resummation, which leads to a smooth
$q_T$ distribution with a shoulder-like behaviour \cite{Catani:1997xc} 
in the vicinity of the back-to-back threshold.

\begin{figure}[htb]
\begin{center}
\begin{tabular}{cc}
\includegraphics[width=0.48\textwidth]{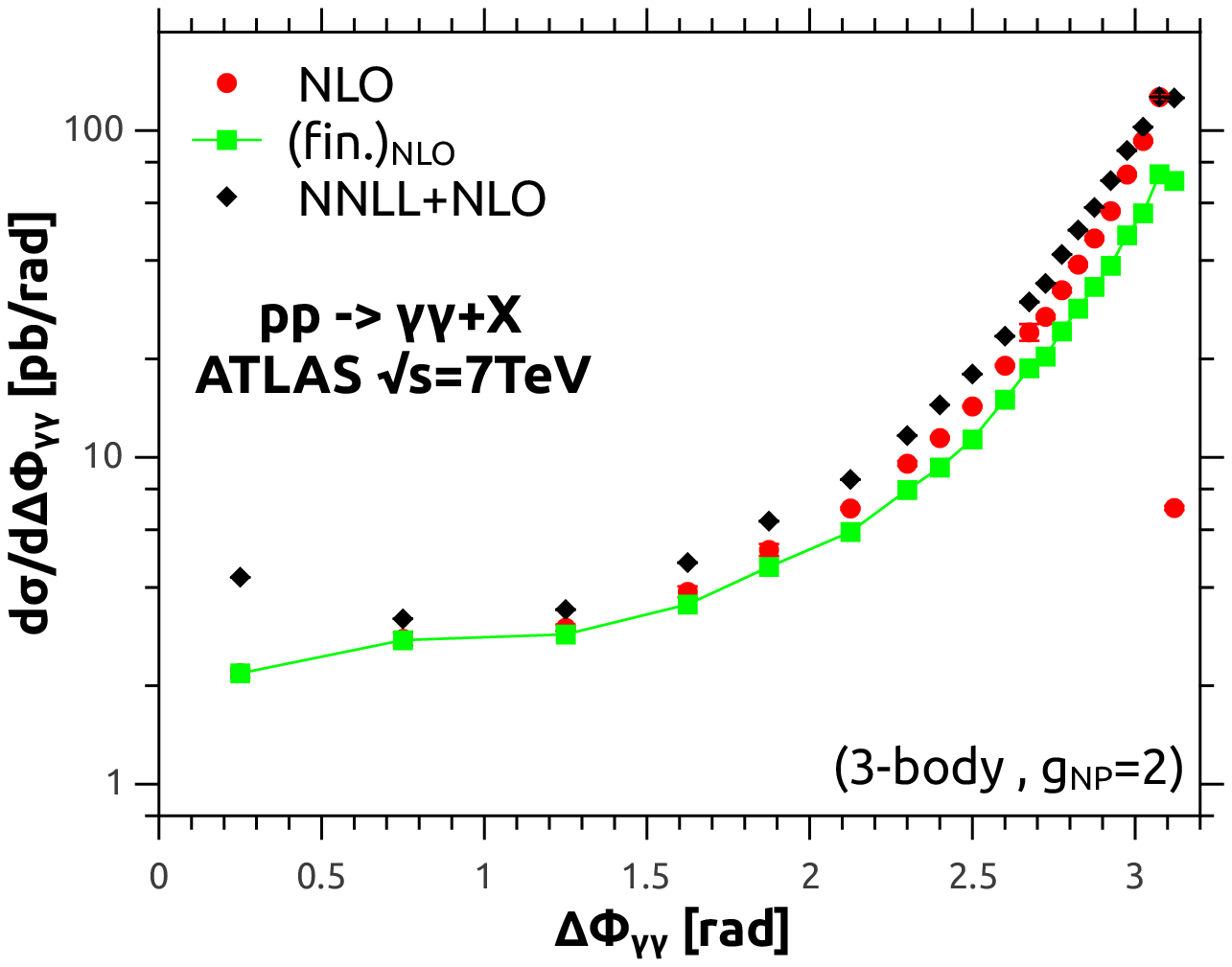} & \includegraphics[width=0.48\textwidth]{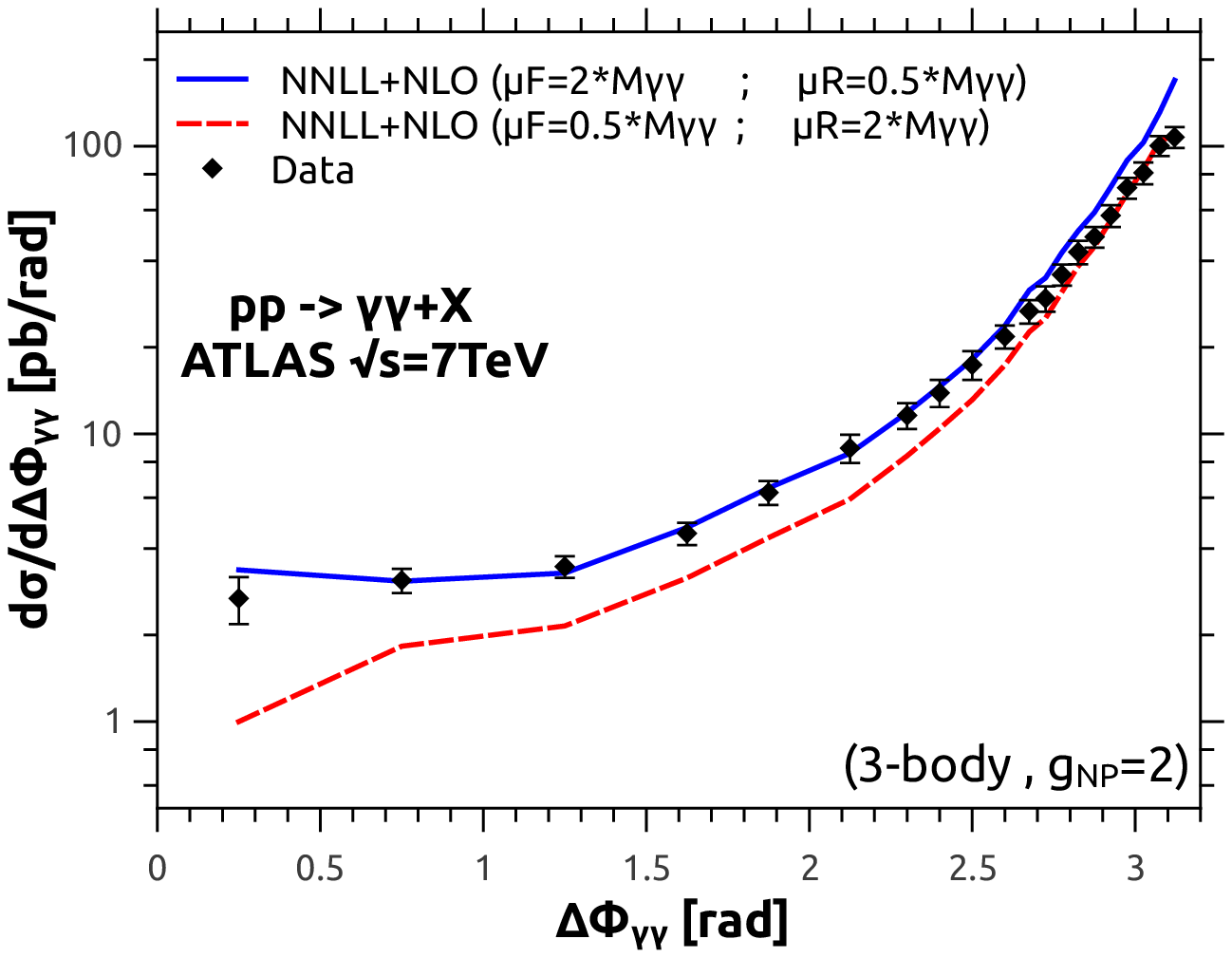}\\
\end{tabular}
\end{center}
\caption{\label{fig6}
{\em The $\Dpgg$ distribution of diphoton pairs at the LHC. In the left panel, we show the fixed order prediction, the complete resummed prediction
and just the resummed contribution at the central scale. In the right, the full NNLL+NLO result (using
the extreme values for $\mu_R$ and $\mu_F$ to estimate the theoretical uncertainty) is compared with the ATLAS data of Ref.~\cite{Aad:2012tba}.}}
\end{figure}

In Fig.~\ref{fig6} we present the results of the cross section as a function of the azimuthal angle $\Dpgg$. In the left panel we compare the fixed order (NLO), finite (NLO) and full (NNLL+NLO) $\Dpgg$ distributions. The fixed order component dominates
the cross section over the whole $\Dpgg$ range.
However, as could be expected, the effect of resummation is stronger for kinematic configurations
near the $\Dpgg\sim\pi$ which correspond to $q_T\sim0$~GeV. As in the case of the fixed order $q_T$ distribution, the $\Dpgg$ fixed
order differential cross section is not well-behaved near the back-to-back configuration: it actually diverges as
$\Dpgg \rightarrow \pi$ ($q_T \rightarrow 0$). The finite contribution (Eq.~\eqref{resfin}) is well-behaved near the back-to-back configuration, and the full result (NNLL+NLO) improves the description in the region near $\Dpgg\sim 0$.

In the right panel of Fig.~\ref{fig6} we compare our theoretical prediction at NNLL+NLO level of 
accuracy with the LHC data~\cite{Aad:2012tba} using the variation
of the $\mu_R$ and $\mu_F$ scales to estimate the theoretical uncertainty. We observe that 
the transverse momentum resummation provides a better description of the data with respect to the 
fixed order result. 
In both panels of Fig.~\ref{fig6} we used the 3-body approach 
to describe the diphoton phase-space. In fact, in this kind of observables 
(also see Fig.~\ref{fig5_1}), the 2-body parametrization again reproduces
the result of the fixed order cross section.

\section{Summary}
\label{sec:summa}

In this paper we performed the transverse momentum resummation for diphoton production at NNLL
accuracy in hadron collisions. 
At small values of $q_T$, the calculation includes the resummation of all logarithmically-enhanced 
perturbative QCD contributions, up to next-to-next-to-leading logarithmic accuracy; 
at intermediate and large values of $q_T$, it combines the resummation
with the fixed next-to-leading order perturbative result. The combination is performed in such a way as to
exactly reproduce the known next-to-next-to-leading order result for the total cross section;
in the end, the calculation consistently includes all perturbative terms up to formal order $\as^2$.
The theoretical uncertainty was estimated by varying the various scales (renormalization, factorization and resummation)
introduced by the formalism as well as the parametrization of the diphoton phase-space.
The result was compared to experimental data, showing good agreement
between theory and experiment over the whole $q_T$ range. With respect
to the fixed-order calculation, the present implementation provides a better description of the data and recovers the correct
physical behaviour in the small $q_T$ region, with the spectrum smoothly going to zero.
The same set-up also allows the calculation of more exclusive observable distributions;
the $q_T$ spectrum of the individual photons and the $\Dpgg$ distribution are given as examples.

\noindent {\bf Acknowledgements.}
We are very grateful to Stefano Catani and Giancarlo Ferrera, for taking a careful look to this paper and making extremely valuable comments about it. LC would like to thank the INFN of Florence for kind hospitality, where parts of the project were carried out. FC would like to thank
the Departamento de F\'\i sica of the Universidad de Buenos Aires and the FCEYN, which supported him during the initial stages of this work.

\end{document}